\documentclass[a4paper,fleqn,usenatbib]{mnras}

\usepackage{newtxtext,newtxmath}
\usepackage[T1]{fontenc}
\usepackage{ae,aecompl}

\usepackage{graphicx}	% Including figure files
\usepackage{amsmath}	% Advanced maths commands
\usepackage{amssymb}	% Extra maths symbols

\usepackage{lscape}
\usepackage{subfigure}
\usepackage{multirow}
\usepackage{url}
\usepackage{threeparttable}
\usepackage{natbib, aas_macros}
\title[X-ray reverberation lags due to AGN winds]{
X-ray reverberation lags of the Fe-K line due to AGN disc winds
}

\author[M.\,Mizumoto et al.]{
Misaki Mizumoto$^{1,2,3}$\thanks{E-mail: misaki.mizumoto@durham.ac.uk, mizumoto.misaki@gmail.com (MM)},
Ken Ebisawa$^{2,3}$,
Masahiro Tsujimoto$^{2}$,
Chris Done$^{1}$,
\newauthor
Kouichi Hagino$^{4}$, \&
Hirokazu Odaka$^{5}$
\\
$^1$Centre for Extragalactic Astronomy, Department of Physics, University of Durham, South Road, Durham DH1 3LE, UK\\
$^2$Institute of Space and Astronautical Science (ISAS), Japan Aerospace Exploration Agency (JAXA), 3-1-1 Yoshinodai, Chuo-ku, Sagamihara,\\
Kanagawa 252-5210, Japan\\
$^3$Department of Astronomy, Graduate School of Science, The University of Tokyo, 7-3-1 Hongo, Bunkyo-ku, Tokyo 113-0033, Japan\\
$^4$Department of Physics, Faculty of Science and Technology, Tokyo University of Science, 2641 Yamazaki, Noda, Chiba 278-8510, Japan\\
$^5$Department of Physics, Graduate School of Science, The University of Tokyo, 7-3-1 Hongo, Bunkyo-ku, Tokyo 113-0033, Japan\\
}

\date{Accepted XXX. Received YYY; in original form ZZZ}

\pubyear{2018}

\begin{document}
\label{firstpage}
\pagerange{\pageref{firstpage}--\pageref{lastpage}}
\maketitle

\begin{abstract}
Short X-ray reverberation lags are seen across a broad Fe-K energy band in more than twenty active galactic nuclei (AGNs).
This broad iron line feature in the lag spectrum is most significant in super-Eddington sources such as Ark 564 ($L/L_{\rm Edd}\sim 1$) and 1H 0707--495 ($L/L_{\rm Edd}\gtrsim 10$). 
The observed lag timescales correspond to very short distances of several $R_g/c$, 
so that they have been used to argue for extremely small `lamp-post' coronae close to the event horizon of rapidly spinning black holes. 
Here we show for the first time that these Fe-K short lags are more likely to arise from scattering in a highly-ionised wind, launched at $\sim 50\,R_g$, rotating and outflowing with a typical velocity of $0.2c$.
We show that this model can simultaneously fit the time-averaged energy spectra and the short-timescale lag-energy spectra of both 1H 0707--495 and Ark 564. 
The Fe-K line in 1H 0707--495 has a strong P-Cygni-like profile, which requires that the wind solid angle is large and that our line of sight intercepts the wind. 
By contrast the lack of an absorption line in the energy spectrum of Ark 564 requires rather face-on geometry,
while the weaker broad Fe-K emission in the energy and lag-energy spectra argue for a smaller solid angle of the wind. 
This is consistent with theoretical predictions that the winds get stronger when the sources are more super-Eddington, supporting the idea of AGN feedback via radiation pressure driven winds.
\end{abstract}


%%%%%%%%%%%%%%%%%%%%%%%%%%%%%%%%%%%%%%%%%%%%%%%%%%

%%%%%%%%%%%%%%%%% BODY OF PAPER %%%%%%%%%%%%%%%%%%

\section{Introduction} \label{sec1}

X-ray illumination of the accretion disc in active galactic nuclei (AGN) produces a fluorescent iron line together with a reflected continuum \citep{geo91,mat91}.  
These scattered photons travel a longer path length than the photons directly observed, producing a 
time delay which is called the reverberation lag.  
This provides a probe of the structure and geometry of the inner accretion flow around the central black hole (e.g. \citealt{utt14}).  

The reverberation lag in AGNs was first seen at the soft energy band ($<1$~keV) in 1H 0707--495 \citep{fab09}.
This was interpreted as the signature of a partially ionised reflector, where the multiple low-energy emission lines are smeared into a pseudo-continuum by extreme relativistic effects (e.g. \citealt{cru06}).
However, this soft X-ray excess region is probably complicated, with additional components from low-temperature Comptonisation \citep{mag98,cze03,nod11b,meh11,jin13,mat14,boi16} 
which can make a soft lead, together with thermalisation of the reprocessed flux that contributes to the soft lag \citep{gar14}. 
Therefore the `soft lag' is not a clean tracer of reverberation.
Instead, the iron line clearly originates in the reflected emission, 
so time lags in the Fe-K band are simpler to interpret. 
These have been reported in multiple AGNs (\citealt{kar16} and reference therein), though are generally less significant than the soft lag.
These Fe-K reverberation lags are seen at high frequencies of $\sim c/100\,R_g$, 
where $R_g=GM/c^2$ is the gravitational radius and $c$ is the speed of light.  
Lag-energy spectra made at this frequency show a broad feature in the 6--9~keV band 
which is lagged behind the adjacent continuum bands by timescale of a few $R_g/c$.

Short-amplitude reverberation lags are often interpreted as evidence for short light travel time lags, indicating reflection from the innermost regions of the accretion disc around a rapidly spinning Kerr black
hole (e.g. \citealt{fab09,kar13b}).  
However, \citet{mil10a} pointed out that the primary component, which has no time delay, contributes a
significant fraction of the flux even in the Fe K band.  
This significantly dilutes the reverberation time delay; 
if the continuum (zero time lag) contributes to half of the photons in the `Fe K' band (with lag time $t_0$), then the observed lag is $t_0/2$.  
Stronger dilution results in shorter lags, and \citet{mil10a,mil10b} and \citet{tur17} showed that the observed lag-frequency plots can instead be explained by more distant scattering clouds orbiting the source at $\sim100\,R_g$. 
\citet{leg12} directly measured the transfer function of Ark 564 and showed that 
scattering from circumnuclear materials at $\sim200\,R_g$ is responsible for the observed time lags.
Winds from the accretion disc are the most likely source of this reprocessing material (see e.g. \citealt{kin03}), 
and winds typically have an outflow velocity which is of order the escape velocity from their launch radii.

\citet{miz18a} (hereafter Paper 1) showed that neutral clouds located at $\sim100~R_g$ with outflow velocity of $0.14c$ (which is the typical value; \citealt{tom11}) can quantitatively explain the frequencies at which the Fe K lags are observed as well as the observed lag amplitudes.  
This also reproduces the width of the broad Fe K feature in the lag-energy spectra as the photons scattered on the near side of the outflow are blueshifted, 
whereas those on the far side are redshifted.  
In the present paper, we extend this analysis by considering a more physically-realistic disc wind picture. 
In fact, Paper 1 assumed that the wind was neutral, whereas any wind launched from the inner disc region should be ionised by the X-ray illumination.
Indeed, ultrafast outflows (UFOs) are now detected as blueshifted absorption lines of highly-ionised (H- or He-like) iron ions in many AGNs (e.g.~\citealt{pou03,ree09,tom10,gof13}).
Paper 1 also assumed that the material was purely radially outflowing, whereas a disc wind should be both rotating and outflowing.
In addition, the wind geometry was assumed to be a section of a hemisphere, 
whereas a bicone geometry is physically more likely \citep{sim08,sim10a,hag15,hag16}.

In this paper, we consider a realistic disc-wind geometry,
in which the outflow gas is rotating, outflowing and highly-ionised. 
We base our calculations on the ionised biconical wind model of \citet{hag15,hag16}, which can fit the UFO features seen in the X-ray energy spectra. 
This uses Monte-Carlo techniques to follow the resonantly-scattered line emission in the highly-ionised gas 
in order to calculate both the absorption and self-consistent emission from the wind. 
Our goal is to construct a model that can explain both the energy spectra and reverberation lags simultaneously. 

In section \ref{sec2}, we explain the method and model in the Monte-Carlo simulation.
Next, we show the resultant energy spectra and lag features in our calculation in section \ref{sec3}.
In section \ref{sec4}, we apply our model to Ark 564 and 1H 0707--495, where the Fe-K reverberation lags are most clearly seen. 
We show that the wind model can explain both the observed time-averaged energy spectra and the broad Fe-K lag in the lag-energy spectra. 
We extend these results into a physically-based picture of radiation powered winds from super-Eddington sources in section \ref{sec5}, and state our conclusions in section \ref{sec6}.

\section{Methods} \label{sec2}

We use the Monte-Carlo simulation code, MONACO (MONte Carlo simulation for Astrophysics and COsmology; \citealt{oda11}), 
which is a general-purpose framework for synthesizing X-ray radiation from astrophysical objects.  
We consider photoionisation, photoexcitation, and line emissions associated with recombination and atomic deexcitation, together with Compton scattering by electrons.
Multiple scatterings and special relativistic effects from the velocity structure in the material are taken into account.
Photons are assumed to be emitted radially from a central point source
(though in practice, this should have a finite extent).  
General relativity effects are ignored as most of the size scales are sufficiently large.  

We assume a biconical wind geometry, as is often used to study radiative transfer in an accretion disc wind (e.g. \citealt{shl93,kni95,sim08,sim10a,hag15,hag16}).  
We adopt the same parameters as used by \citet{hag16} to explain the spectral features seen in 1H 0707--495, i.e., 
the wind is assumed to fill an axisymmetric bicone between $45^\circ$ to $56.^\circ3$ 
(so that the solid angle from the focal point $d$ below the black hole is $\Omega/2\pi=0.15$), 
launched at $d=R_{\rm min}=50\,R_g$, corresponding to an escape velocity of $0.2c$.  

The radial velocity follows an extension of the CAK velocity law \citep{cas75}, i.e.,
\begin{equation}
v_r(l)=v_0+(v_\infty-v_0)\left(1-\frac{R_{\rm min}}{R_{\rm min}+l}\right)^\beta, \label{eq:vel}
\end{equation}
where $v_r(l)$ is a radial velocity of the wind along a streamline of length $l$ from its launch point,
$v_0=10,000$~km/s is the initial velocity (same as the turbulent velocity),
$v_\infty=0.2c$ is the terminal velocity, and $\beta=1$ is the acceleration index.
The rotational velocity is determined by conservation of the angular momentum.
The density $n(r)$ depends on the mass loss rate in the wind, $\dot{M}_{\rm wind}$, as 
\begin{equation}
\dot{M}_{\rm wind}=\mu m_pn(r)v(r)4\pi D^2 (\Omega/4\pi), \label{eq:Mdot}
\end{equation}
where $\mu$ is the mean molecular weight, $m_p$ is the proton mass, and $D$ is the distance from the focal point.
We assume that $\dot{M}_{\rm wind}/\dot{M}_{\rm Edd}=0.2$ and calculate $n(r)$, where the
geometry is self-similar, so that 
the ionisation degree and the column density are not dependent on the BH mass \citep{hag16}.
Here we assume that $M_{\rm BH}=2\times10^6\,M_\odot$. i.e. $1R_g=10$~s.

The calculation grid is made by splitting the bicone into the 100(radial) $\times$64(azimuthal) $\times$2(polar) cells.  
The ion populations are calculated in each radial shell by using XSTAR (version 2.39; \citealt{kal04}).  
The central source is assumed to emit a power-law spectrum with photon index $\Gamma=2.6$ and ionising luminosity $L_{\rm X}=0.003L_{\rm Edd}$, so that 
the ionisation structure can be calculated in each grid point from $\xi=L_{\rm X}/n(r)r^2$ 
(where $L$ is calculated in the co-moving wind frame taking account of special relativistic shifts). 
The resultant ion populations in each grid point are calculated sequentially from the inner to the outer shells.  
Typical electron temperature and ionisation degree are $\sim10^5$~K and $\log\xi\sim 5$, respectively,
so the wind consists of only highly-ionised material.  
Therefore we consider transitions of only H- and He-like ions of Fe and Ni.

\section{Results} \label{sec3}

\subsection{Spectra from the wind}

\begin{figure}
\begin{center}
\includegraphics[width=2.5in,angle=270]{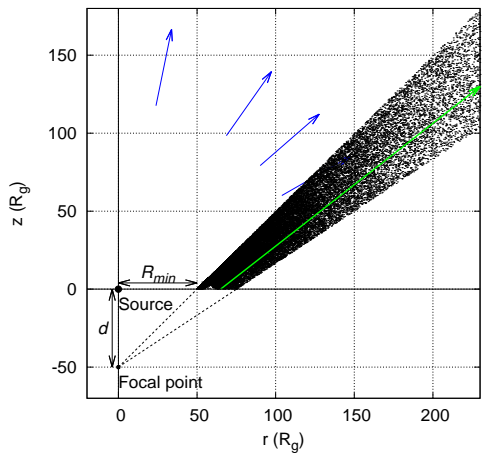}
\caption{
Positions of the last scattering for 1 in every 1000 photons in the simulation. 
The density of points is highest on the inner face of the wind, close to the launch radius, 
indicating that the wind is optically thick across its base. 
The density drops at larger radii, as the wind velocity increases along the streamline (green).
The four blue arrows show the inclination angles studied in this paper.}
\label{fig:geometry}
\end{center}
\end{figure}

\begin{figure}
\centering
\includegraphics[width=2.3in,angle=270]{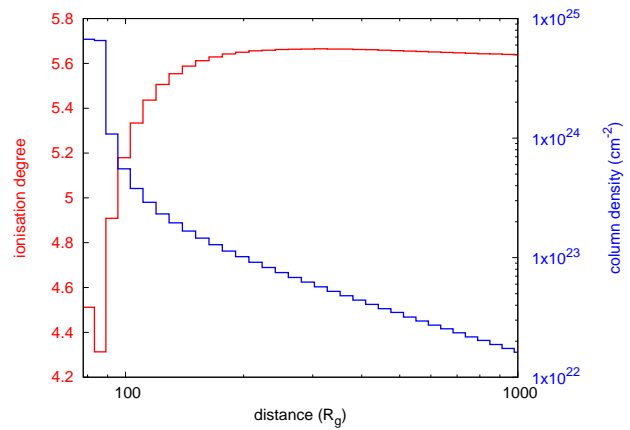}
\caption{
Ionisation degree (red, left axis) and column density (blue, right axis) along the streamline 
(the green arrow in Fig.~\ref{fig:geometry}).
The horizontal axis shows the distance from the focal point.
Each bin corresponds to a radial shell in the XSTAR calculation.
}
\label{fig:r-dependence}
\end{figure}

The wind geometry is shown in Fig.~\ref{fig:geometry}, where the points show locations of the last scattering for 0.1\% of all the photons in the simulation.  
The scattered photon density is largest at the upper face of the wind at fairly small radii, 
showing that the wind is optically thick across its base. 
Most of the scatterings take place in the inner $\lesssim100~R_g$. 
The green line shows a streamline along the wind, and Fig.~\ref{fig:r-dependence} shows the ionisation parameter (red, left axis) and column density (blue, right axis) along this line.  
The wind velocity along the streamline is relatively slow close to the launch point, 
so the velocity is slowest and the density is highest at the base. 
Thus the region within $100\,R_g$ has the lowest velocities, lowest $\xi$ and highest $N_{\rm H}$.
H- and He-like iron ions are dominant for our assumed parameters, and
the wind efficiently produces the associated resonance iron emission lines.
  
The wind should respond in ionisation state more or less simultaneously with changes in the continuum at these densities. 
Such ionisation changes are seen in IRAS 13224--3809 \citep{par17b}, a `twin' of 1H 0707--495 due to the similarity in their overall properties \citep{lei04,pon10}.
In this paper, however, we neglect such possible time variability of the ionisation state in order to focus on overall properties of the scattering. 

\begin{figure*}
\begin{center}
  \subfigure{
    \resizebox{8cm}{!}{\includegraphics[width=65mm,angle=270]{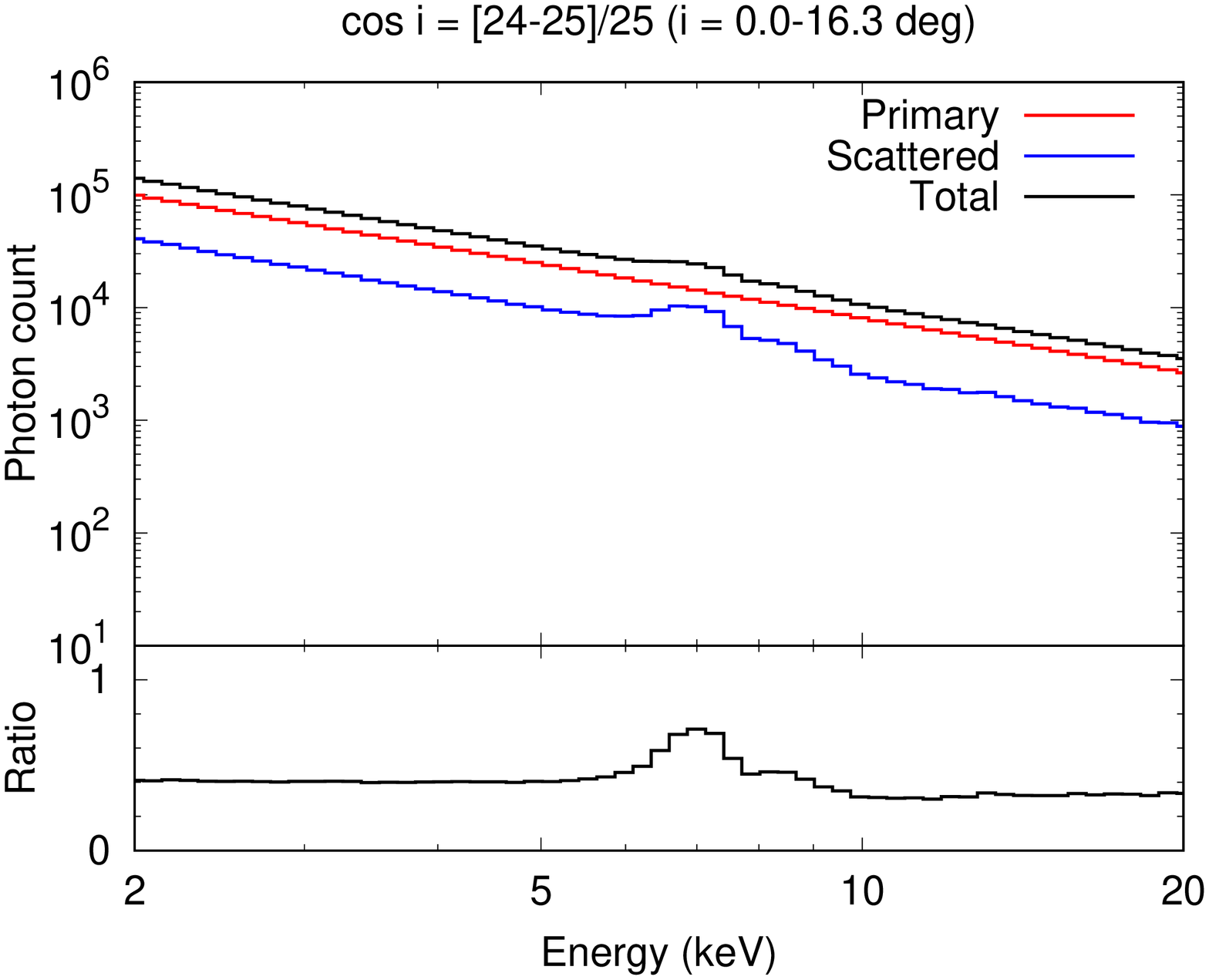}}
    \resizebox{8cm}{!}{\includegraphics[width=65mm,angle=270]{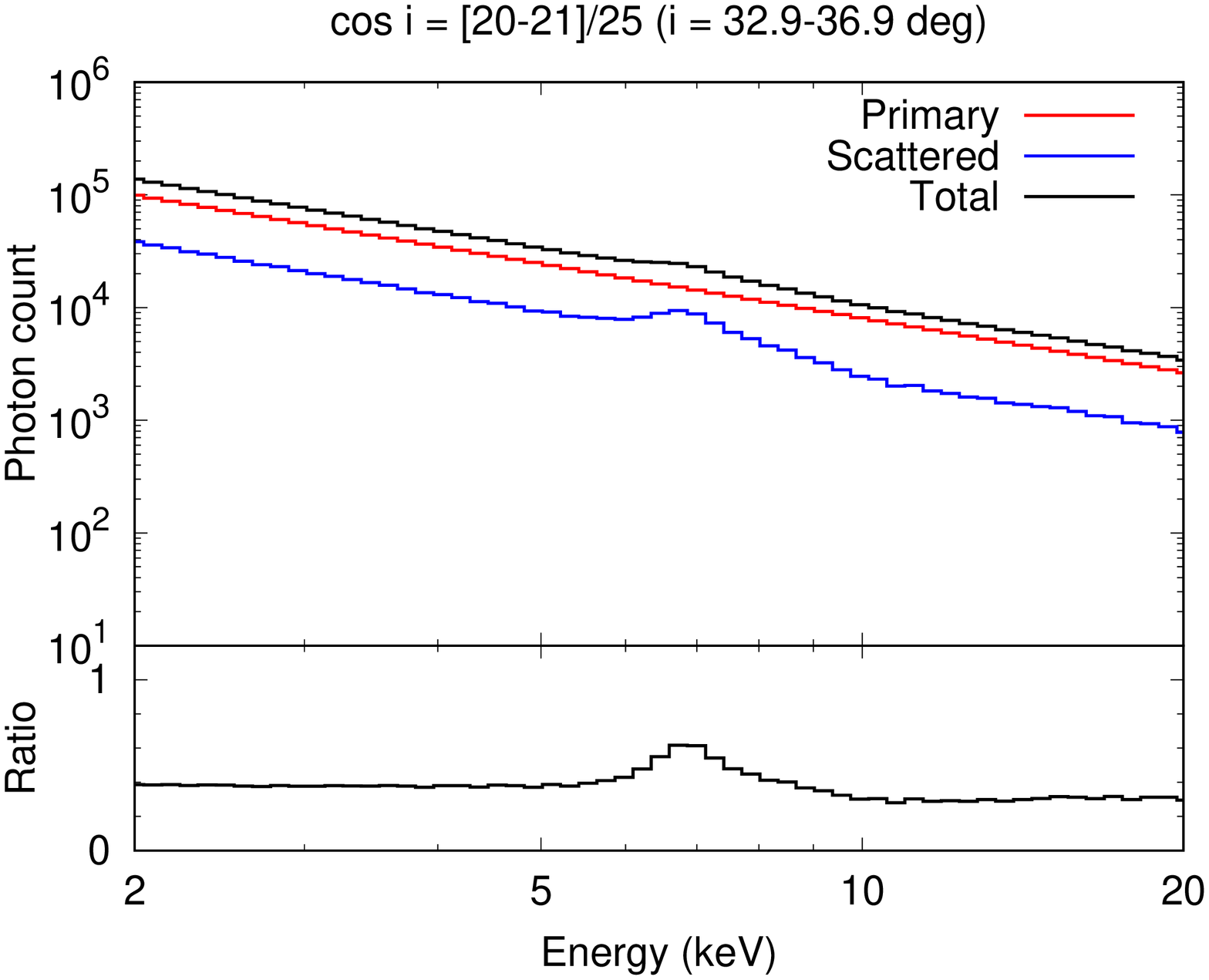}}
  } \subfigure{
    \resizebox{8cm}{!}{\includegraphics[width=65mm,angle=270]{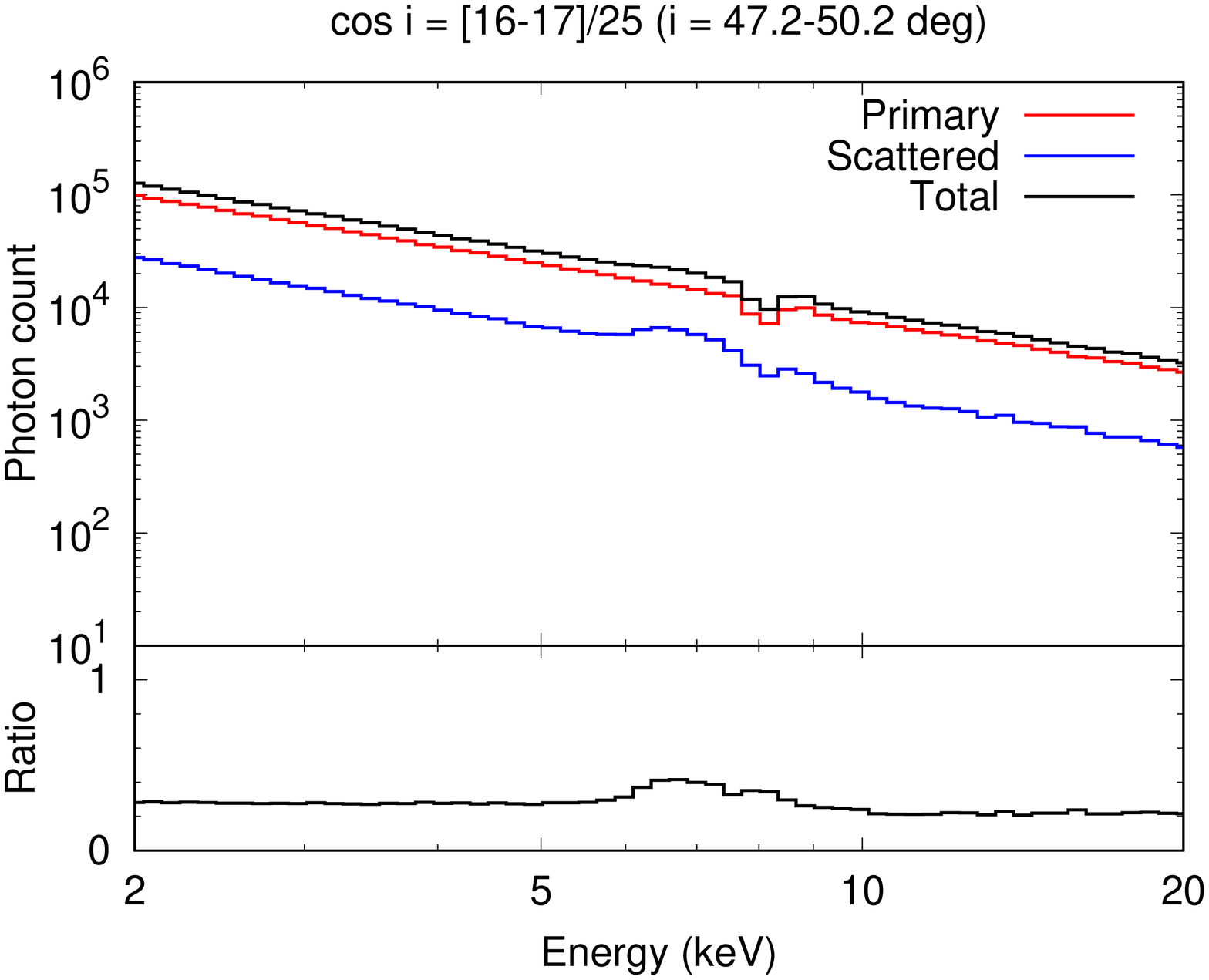}}
    \resizebox{8cm}{!}{\includegraphics[width=65mm,angle=270]{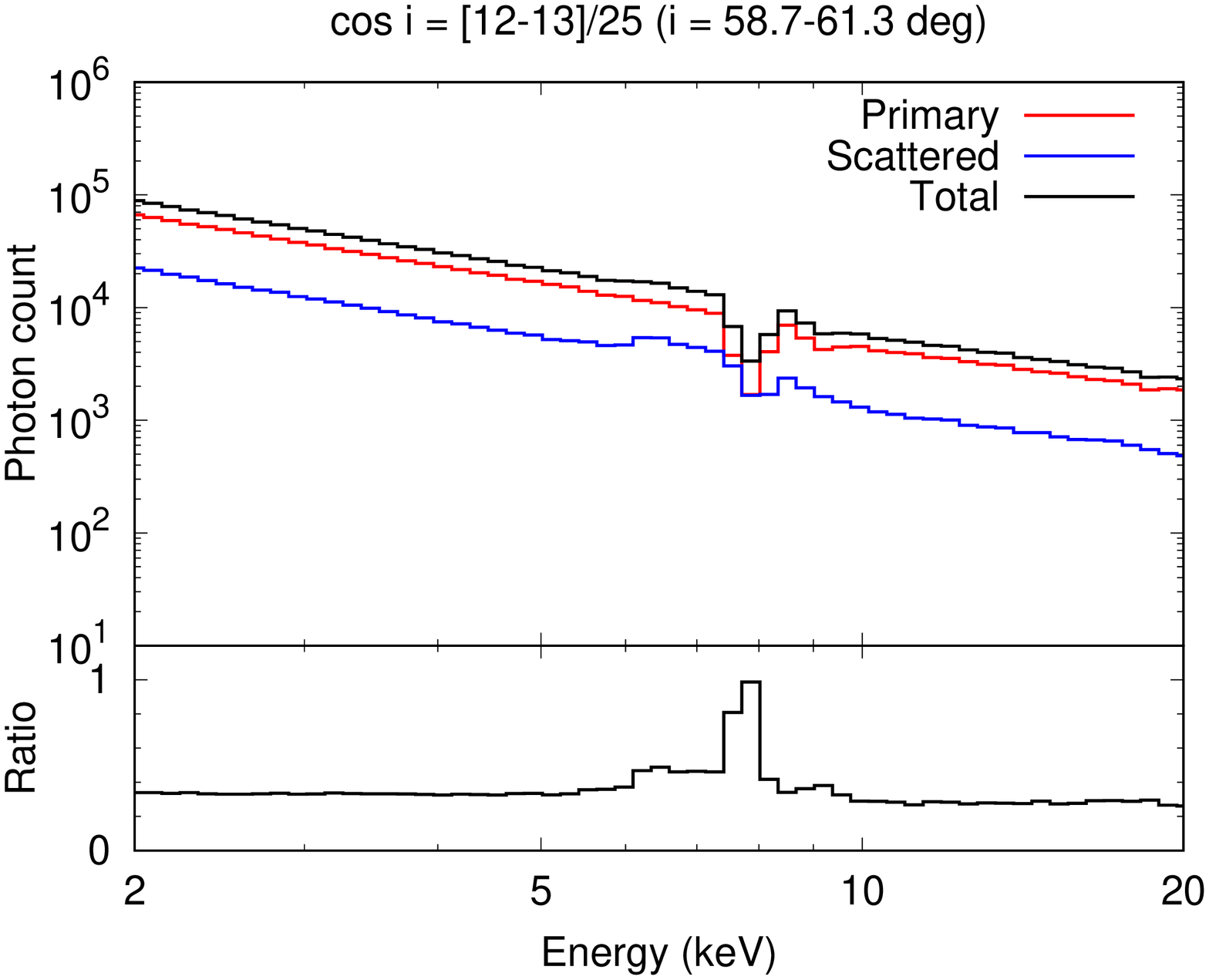}}
  }\caption{
  Energy spectra of primary (red), scattered (blue), and total (black) components expected in the wind model.  
  Those for four representative inclination angles are shown. 
  Lines of sight of the first two do not intercept the wind, so these spectra do not show strong absorption line features, 
  while the latter two cut across the wind and the corresponding spectra show the progressively stronger absorption lines which arise from the progressively larger optical depth of material in the line of sight. 
    The bottom panels show ratios of the scattered component to the primary spectrum.}
\label{fig:spectrum_wind}
\end{center}
\end{figure*}

\begin{figure}
\begin{center}
\includegraphics[width=2.3in,angle=270]{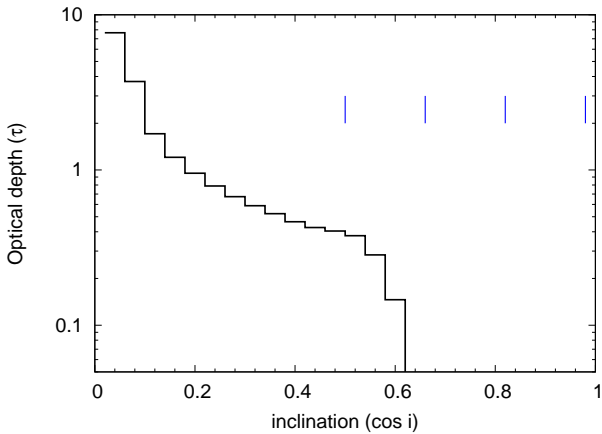}
\caption{ 
The electron scattering optical depth in the wind for different inclination angles.  
The vertical blue lines correspond to the inclination angles shown in Fig.~\ref{fig:geometry}.
 In the face-on case ($\cos i>17/25$) there is no wind in the line of sight, so the optical depth is zero. 
 }  
\label{fig:tau}
\end{center}
\end{figure}

We collect energy spectra in 24 bins in $\cos i$.  
Fig.~\ref{fig:spectrum_wind} shows these for the four representative lines of sight marked on Fig.~\ref{fig:geometry}, 
going from face on ($\cos i=(24-25)/25$, i.e. $\sim 11^\circ$),
to moderately inclined but not intercepting the wind ($\cos i=(20-21)/25$; $35^\circ$), 
to grazing the top part of the wind ($\cos i=(16-17)/25$; $49^\circ$), 
and looking deeper into the wind ($\cos i=(12-13)/25$; i.e. $60^\circ$). 
The primary flux (red) is simply the intrinsic flux for directions which do not intercept the wind, 
but is reduced by $e^{-\tau_{\rm e}}$ (where $\tau_{\rm e}$ is the optical depth of the gas) for directions which intercept the wind (Fig.~\ref{fig:spectrum_wind}). 
The optical depth through the wind for the highest inclination angle considered here is $\sim 0.4$ (Fig.~\ref{fig:tau}), so the continuum is reduced by a factor $0.67$. 

There is absorption in the wind as well as electron scattering, so absorption lines are seen on the primary emission where the line of sight intercepts the wind.  
The optical depth in the wind material increases strongly at larger inclination angles 
(see Fig.~\ref{fig:tau}) so the absorption line equivalent width increases with the inclination angle. 
The intrinsic width of the absorption line depends on the velocity gradient along the line of sight.  
This is always large enough to mix the H- and He-like lines (6.7~keV and 7.0~keV in the rest frame, respectively) together, 
so only a single absorption feature is seen (see also \citealt{hag16}). 
Higher inclination angles cut across closer to the base of the wind, where the outflow is slower, 
so the degree of blueshift decreases. 

The scattered continuum (blue) instead shows a broad emission line from 6--7~keV, together with a broad dip at 8.6--10~keV corresponding to the ionised edges (Fig.~\ref{fig:spectrum_wind}). 
The scattered flux is also absorbed by the wind at high inclination angles, so
there is an absorption line in the scattered emission along lines of sight intercepting the wind. 
The emission line in the scattered flux is always broader than the absorption line, 
since it is produced by the winds over all azimuths, sampling a wider range of the projected velocities. 
The black lines show the total continuum (primary plus scattered) along each line of sight. 
The lower panel on each spectrum in Fig.~\ref{fig:spectrum_wind} show
the ratio of the scattered emission to the primary one at each inclination angle;
we stress that this is the key factor which determines the dilution in the lag-energy spectra (see equation \ref{eq:dilutionE}). 

\subsection{lags from the wind}
\begin{figure}
\begin{center}
\includegraphics[width=2.3in,angle=270]{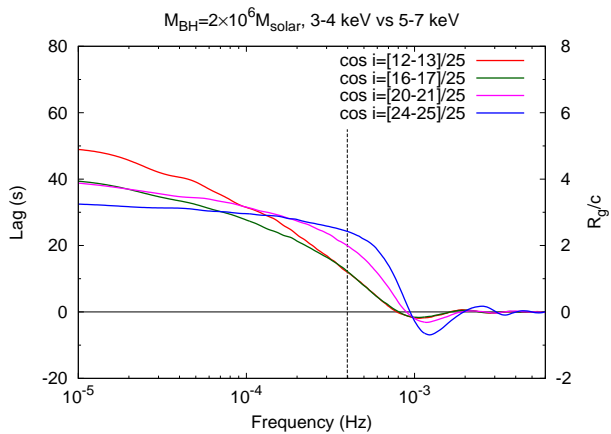}
\caption{
Lag-frequency plots in the wind model for different inclination angles. 
Positive means hard lags, i.e., photons in 5--7 keV lag behind those in 3--4~keV. 
The vertical line indicates $10^{-3.4}$ Hz, which is used to compare Fe-K lag in different conditions below.
}
\label{fig:lagvsf_wind}
\end{center}
\end{figure}

We calculated the time lags predicted in this model between a soft (3--4~keV; `power law dominated') band and a hard (5--7~keV; `Fe line dominated') band.  
These are extracted from the Monte-Carlo simulation as described in Paper 1.  
Fig.~\ref{fig:lagvsf_wind} shows the resulting lag-frequency plot for different inclination angles.  
These are fairly similar, showing a hard (Fe line) lag of around 50~s at the low-frequency limit, corresponding to $5\,R_g/c$ which is much smaller than the distance of the region where iron lines are typically produced ($\sim 100\,R_g$).
This is due to the presence of the continuum photons with no lags in the `Fe line dominated' band, 
which dilutes the true iron line lag.  
The frequency at which the lag drops towards zero is a reliable measure of the distance of the iron line region, with $\sim2-5\times10^{-4}$~Hz, 
which corresponds to $c/500\,R_g - c/200\,R_g$ (see also paper 1).

The small negative (i.e. soft) lags seen above $\sim10^{-3}$~Hz are mathematical artefacts of the Fourier transforms (see also \citealt{mil10a} and Paper 1).  
Paper 1 analytically shows that these negative lags emerge from phase wrapping; 
a delta function response with time delay of $\Delta\tau$ or a top-hat function spanning delay times of
$0-2\Delta\tau$ will both show negative lags above frequencies of $\sim1/(2\Delta\tau)$.

\begin{figure}
\begin{center}
\includegraphics[width=2.2in,angle=270]{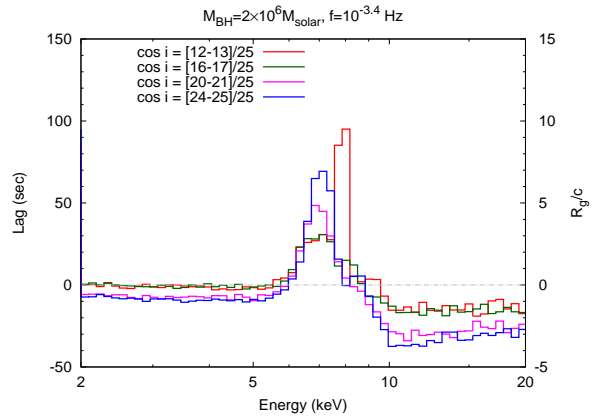}
\caption{
Lag-energy plots in the wind model for different inclination angles at $10^{-3.4}$~Hz ($=c/250\,R_g$).
}
\label{fig:lagvsE_wind}
\end{center}
\end{figure}

We choose $10^{-3.4}=4\times 10^{-4}$~Hz, which corresponds to $c/250 R_g$, 
as the highest frequency at which the reverberation lags are clearly seen.  
This frequency will be the best to study the Fe-K line lags in real data, because 
propagation lags due to fluctuations through the accretion flow (e.g. \citealt{are06}) are less affected.  
We use the Monte-Carlo simulation to extract the lag-energy spectra 
(using a reference band from 2--10~keV, not including the energy band of interest) 
at this frequency over our four inclination angle bins (Fig.~\ref{fig:lagvsE_wind}).  
There is clearly an increase in lag over a broad range in energies centred around 7~keV.  
This is due to the lagged iron emission line, which is both blue and redshifted 
due to stratification of the projected line of sight velocities 
from the rotating/outflowing winds over all the azimuths and all the radii of the winds.
Amplitudes of the broad lag feature centring at $\sim7$~keV decreases from 70~s to 30~s 
with increasing the inclination angle 
as the broad emission line is stronger in the face on spectra (Fig.~\ref{fig:spectrum_wind}).
At the highest inclination angle, a prominent sharp feature at 8~keV is superimposed, 
corresponding to the flux decrease in the primary component 
at the absorption line energy (Fig.~\ref{fig:spectrum_wind}).

We use the formalism of Paper 1 to estimate the lag amplitudes as a function of energy, $\tau(E)$.  
In the low frequency limit these asymptote to
\begin{equation}
\Delta\tau(E)_{f\to 0} \simeq \left(\frac{R(E)}{P(E)} - \frac{R_{\rm tot}}{P_{\rm tot}}\right)\Delta \tau
\label{eq:dilutionE}
\end{equation}
where $P(E)$ and $R(E)$ denote the primary and scattered contribution at energy $E$, respectively (see the bottom panels of Fig.~\ref{fig:spectrum_wind}), 
the suffix ``tot'' means the total energy band (2--10~keV), and 
$\Delta\tau$ is the mean intrinsic lag of $100\,R_g/c$ (see equation 12 in Paper 1).  
This means that the lag amplitude becomes larger in the energy band where the scattered spectrum is stronger and/or the primary spectrum is weaker.  
Fig.~\ref{fig:spectrum_wind} shows that the scattered spectra have the broad iron line emission which peaks at $\sim7$~keV.
Their equivalent widths increase by a factor 2 as the inclination angles decrease because the flux of the scattered spectra relatively increases.
These broad features are very similar to the results of Paper 1, but the lag amplitudes are larger because the resonance scattering is more efficient than the neutral fluorescence.
In addition to it, in the highest inclination angle case there is a deep absorption line around 8~keV 
in the primary component, resulting in a larger fraction of the scattered contribution and a sharp increase in lag at this narrow band (Fig.~\ref{fig:lagvsE_wind}).  

\begin{figure*}
\begin{center}
\subfigure{
\resizebox{8cm}{!}{\includegraphics[width=65mm,angle=270]{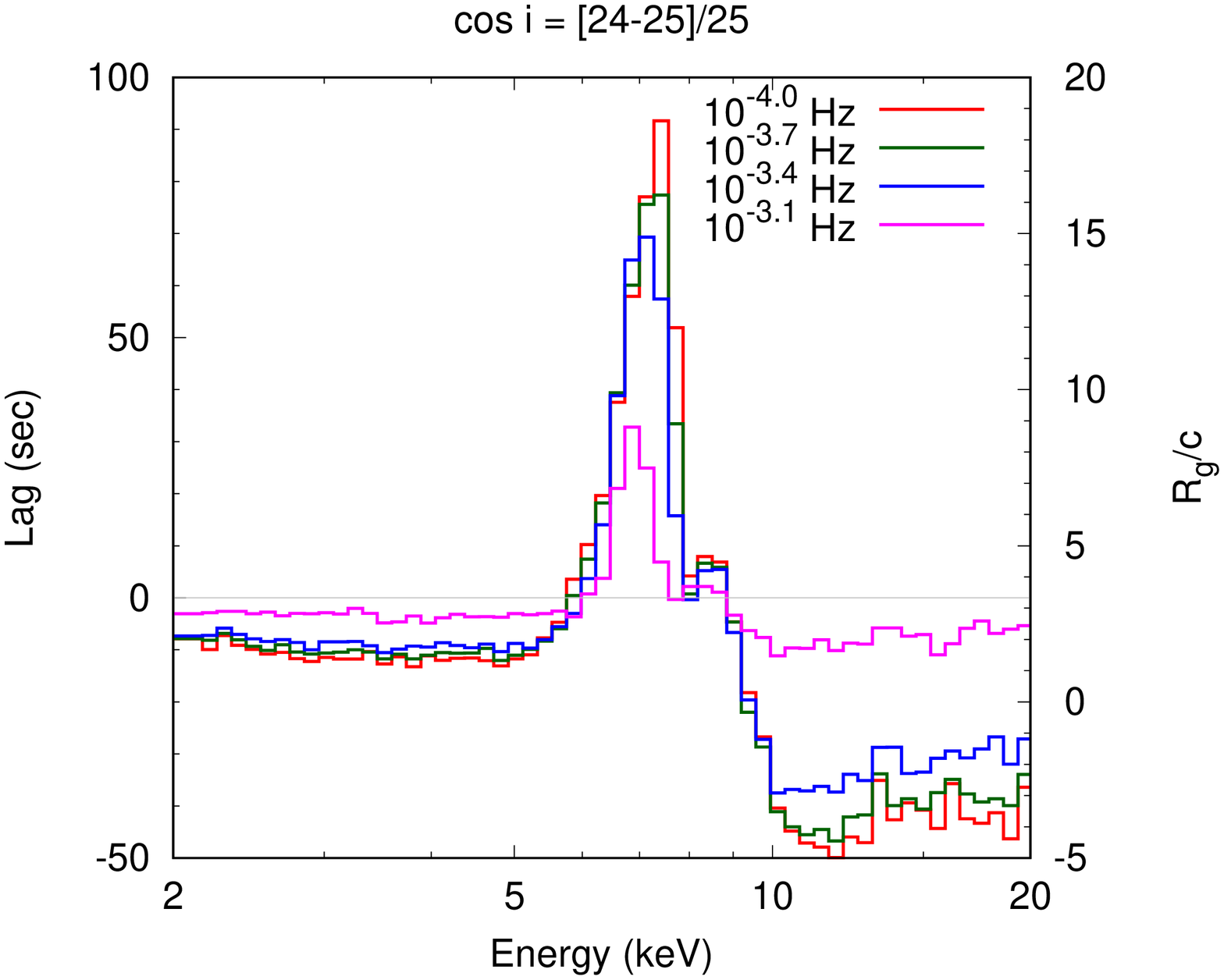}}
\resizebox{8cm}{!}{\includegraphics[width=65mm,angle=270]{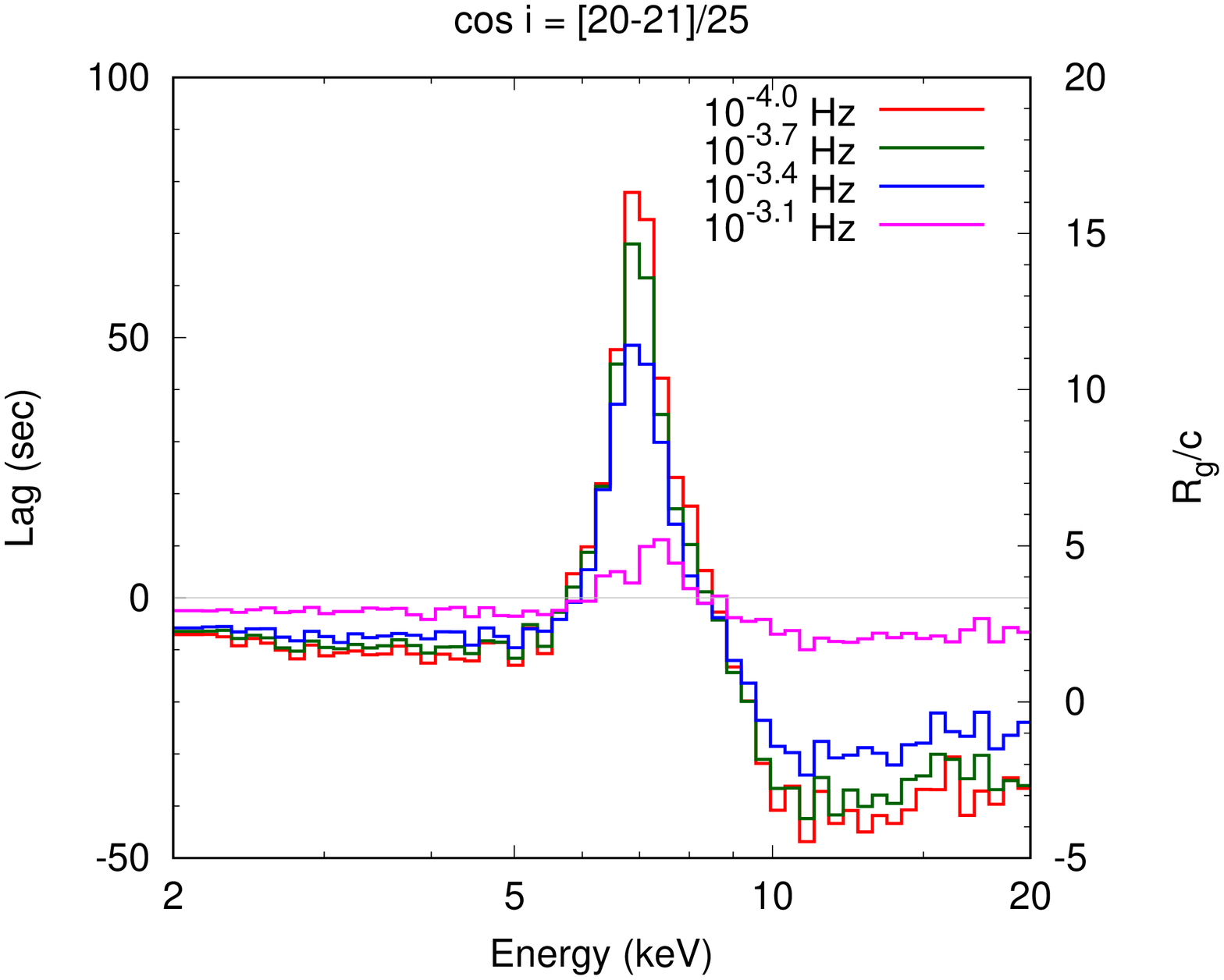}}
}
\subfigure{
\resizebox{8cm}{!}{\includegraphics[width=65mm,angle=270]{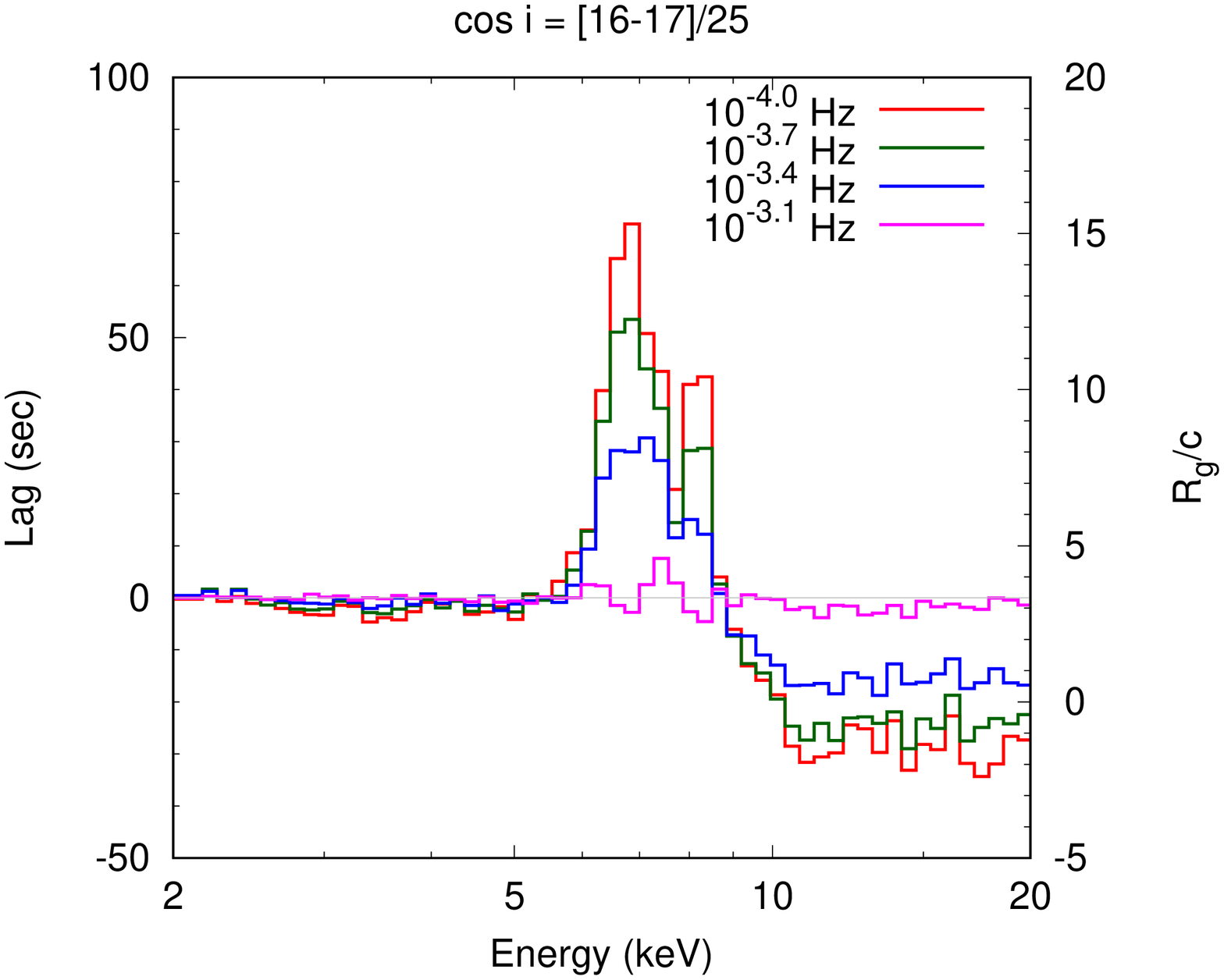}}
\resizebox{8cm}{!}{\includegraphics[width=65mm,angle=270]{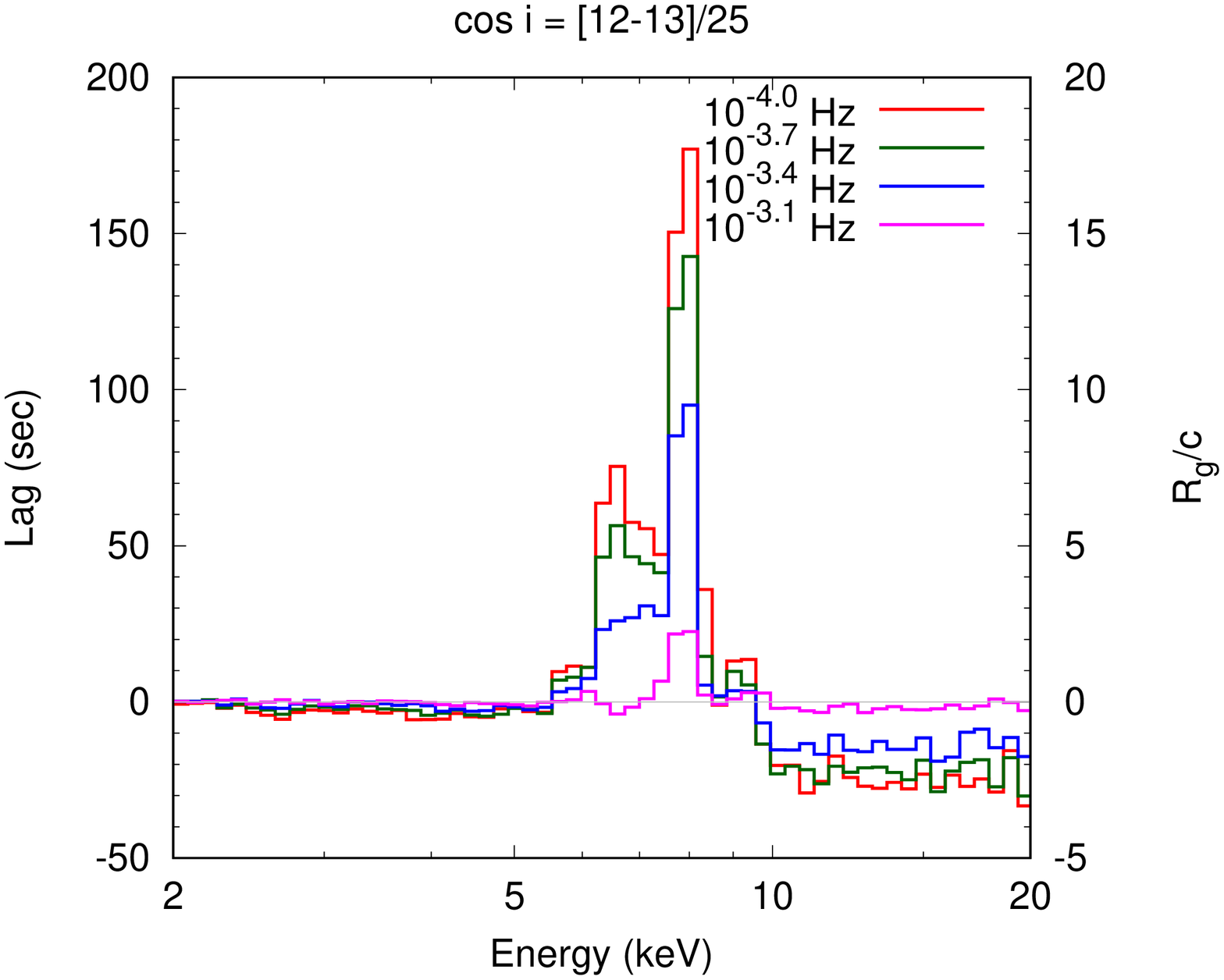}}
}
\caption{
Lag-energy plots in the wind model for different frequencies and the four representative inclination angles}
\label{fig:freqdepend}
\end{center}
\end{figure*}

We explore the frequency dependence of the lag-energy spectra in Fig.~\ref{fig:freqdepend}.  
The blue line shows the lag-energy at $10^{-3.4}$~Hz, which is the same as in Fig.~\ref{fig:lagvsE_wind}.  
At higher frequencies ($10^{-3.1}$~Hz, magenta lines), the variability is too fast to probe the wind, 
so there are few lagged signals.  
As the frequencies drop ($10^{-3.7}$~Hz in green lines and $10^{-4}$~Hz in red lines), 
the distance that the lower frequency may probe becomes larger, so the Fe-K lag gets larger.  
When the inclination angle is large enough (the bottom two panels in Fig. \ref{fig:freqdepend}), 
photons in the red side of the broad emission lines, which come from the far side of the wind, 
travel longer paths than those in the blue side.  
Therefore the lag features in the red side get larger at the lower frequencies (also see Fig.~7 in Paper 1).
Also, the sharp blue peak due to the absorption line gets stronger as the frequency decreases.
In these manners, the Fe-K lag features both in the energy and frequency domains reflect the particular configuration of the system, 
such as the inclination angles and distribution of the matter around the central source.

\subsection{Disc reflection}

Our wind is launched at a radius of $50\,R_g$, so there should be a disc inside.  
Below, we estimate possible effects of the disc reflection in the same Monte-Carlo framework.
We assume that the disc is neutral, optically-thick, geometrically-thin ($h/r=0.001$, 
where $h$ is the thickness), and in Keplerian motion.  
We envisage a finite source with a size of $\sim5-10\,R_g$, and the inner edge of the reflecting disc to be $10\,R_g$.
However, for computational simplicity, we approximate this extended illuminator by a  point-like 
``lamp post'' corona located at a height of $5\,R_g$ (see also \citealt{gar17}).  
General relativistic effects are not included.  
The solid angle of the inner disc  ($10-50\,R_g$) seen by the source is $\Omega/2\pi=0.35$.  
This is smaller than the solid angle subtended by the optically thickest part of the wind ($r<100\,R_g$ and $z<50\, R_g$ in Fig.~\ref{fig:geometry}), which is $\Omega/2\pi>0.55$. 
The larger solid angle for the wind means that scattering flux from the wind should always dominate over the reflected flux even if the disc is somewhat ionised. 

\begin{figure*}
\begin{center}
\subfigure{
\resizebox{8.5cm}{!}{\includegraphics[width=3.2in,angle=270]{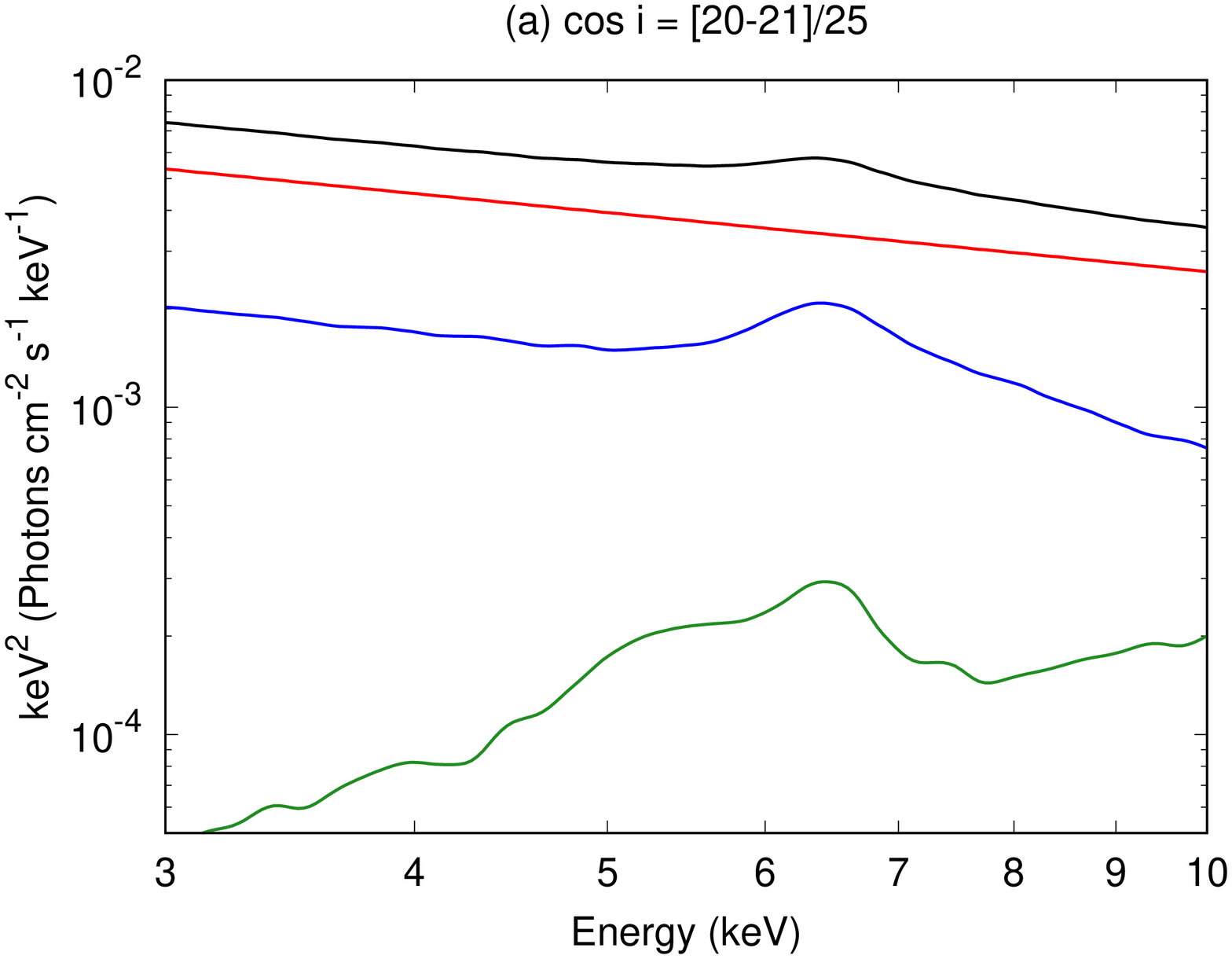}}
\resizebox{8.5cm}{!}{\includegraphics[width=3.2in,angle=270]{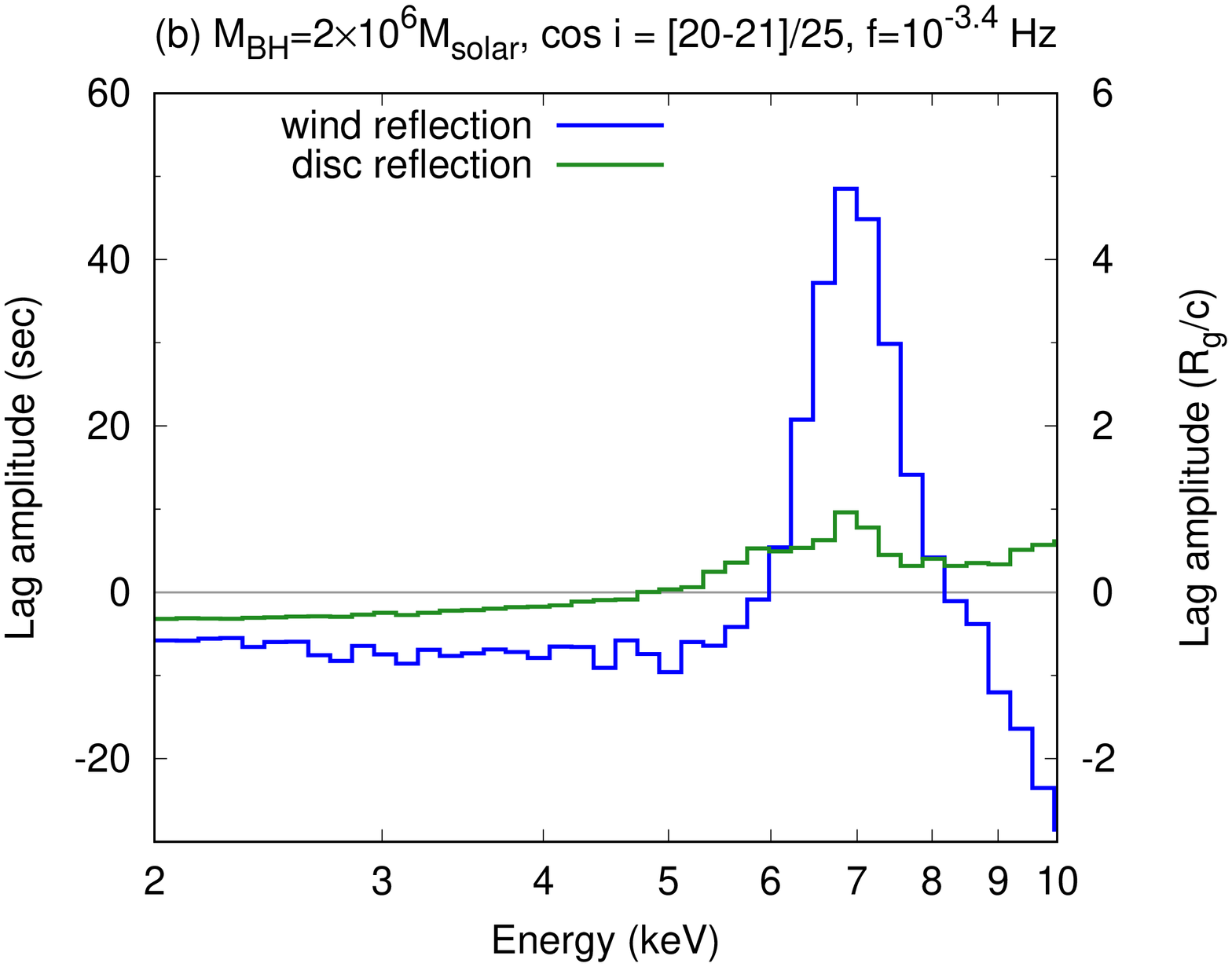}}
}
\caption{
(a) Simulated energy spectra for the combined disc plus wind geometry for the face-on case. 
The red line shows the primary emission, while the blue shows the wind-scattered component 
as in Fig.~\ref{fig:spectrum_wind}a. 
The green line shows the additional reflection component from the inner disc of $10-50R_g$.
The black line shows the total spectrum (primary+wind-scattered+disc-reflection). 
(b) Lag-energy plots at the frequency of $c/250\,R_g$. 
The lags from the wind reflection are always larger than those from the  disc reflection, because 
the wind contributes much more to the scattered emission due to combination of the larger solid angle and the higher ionisation state.
}
\label{fig:winddisc}
\end{center}
\end{figure*}

Figure \ref{fig:winddisc}(a) shows the simulated spectra for the combined inner disc and wind geometry.  
The Fe-K fluorescence line from the inner disc at 6.4~keV is broadened by the fast rotational velocity,
and the double peak profile is seen.  
While this is not completely negligible, the scattered emission from the wind is always larger than the disc-reflected emission, 
mainly due to the larger solid angle of the wind and 
secondly as the disc material is less ionised than the wind material (e.g. \citealt{bal17}) and 
the resonance scattering in the ionised wind can more effectively produce emission lines than from the disc.  
Figure \ref{fig:winddisc}(b) shows the lag-energy spectrum of the combined disc(green)-wind(blue) geometry at the frequency of $c/250\,R_g$.  
Again, there is only a small contribution from the inner disc. 
The majority of the broad iron line in the lag-energy spectra at this frequency comes from the wind.

\section{Comparison with observations}\label{sec4}

We apply our model to observations where the lag-energy spectra reveal a broad feature around the iron line.  
We use X-ray archival data of {\it XMM-Newton} \citep{jan01}, because the EPIC-pn has a large effective area \citep{str01}, which brings us good photon statistics.
In addition, {\it XMM-Newton} has a higher orbit and thus longer orbital period ($\sim48$~hr) than other satellites such as {\it Suzaku} ($\sim1.6$~hr), 
so that long contiguous datasets can be obtained, which is essential to calculate time lags.  
From the compilation by \cite{kar16}, we choose Ark 564 and 1H 0707--495 where the reverberation lags appear best constrained by the data.  
We point out that these two objects have similar features in the lag-energy spectra, but very different time-averaged spectra \citep{kar13,kar13b}; we aim to understand these similarities and differences in the context of our wind-scattering model.
The data are reduced using the XMM-Newton Software Analysis System ({\tt SAS}, v.16.1.0) and the latest calibration files. 
We use all the archival data in XMM-Newton with the exposure time (after removing high background periods) longer than 10~ks (Tab.\ \ref{tab:log}).
The source spectra are extracted from a circular region of a radius of $30^{\prime\prime}$ centred on the source,
whereas the background spectra are from a circular region of a radius of $45^{\prime\prime}$ in the same CCD chip near the source without chip edges nor serendipitous sources.

\begin{table*}
	\centering
	\caption{Observation log}
	\label{tab:log}
	\begin{tabular}{cccccc}
		\hline
&Ark 564 &&&1H 0707--405 &\\
\hline
ID & Exposure (s) & Countrate (s$^{-1}$) & ID & Exposure (s) & Countrate (s$^{-1}$)\\
		\hline
0206400101 &69150	&36.7&0110890201	&37824	&1.3\\
0670130201 &41392	&59.0&0148010301	&68095	&4.2\\
0670130301&38807&39.3&0506200201	&26861	&0.8\\
0670130401&38471&40.1&0506200301	&35826	&2.2\\
0670130501&46845&47.5&0506200401	&14655	&4.4\\
0670130601&41668&41.2&0506200501	&32473	&6.0\\
0670130701&32341&26.6&0511580101	&99628	&3.7\\
0670130801&40496&41.4&0511580201	&66365	&5.6\\
0670130901&38678&54.7&0511580301	&59821	&4.9\\
&&&0511580401	&66588&4.0\\
&&&0554710801	&64524&0.3\\
&&&0653510301	&103604&4.3\\
&&&0653510401	&102075&5.6\\
&&&0653510501	&95740&4.2\\
&&&0653510601	&97621&5.2\\
\hline
(total/average) & 387848 & 42.9 & (total/average)&971700 &3.8 \\
	\hline
	\end{tabular}
\end{table*}

\begin{figure*}
\begin{center}
\subfigure{
\resizebox{8.5cm}{!}{\includegraphics[width=2.7in,angle=270]{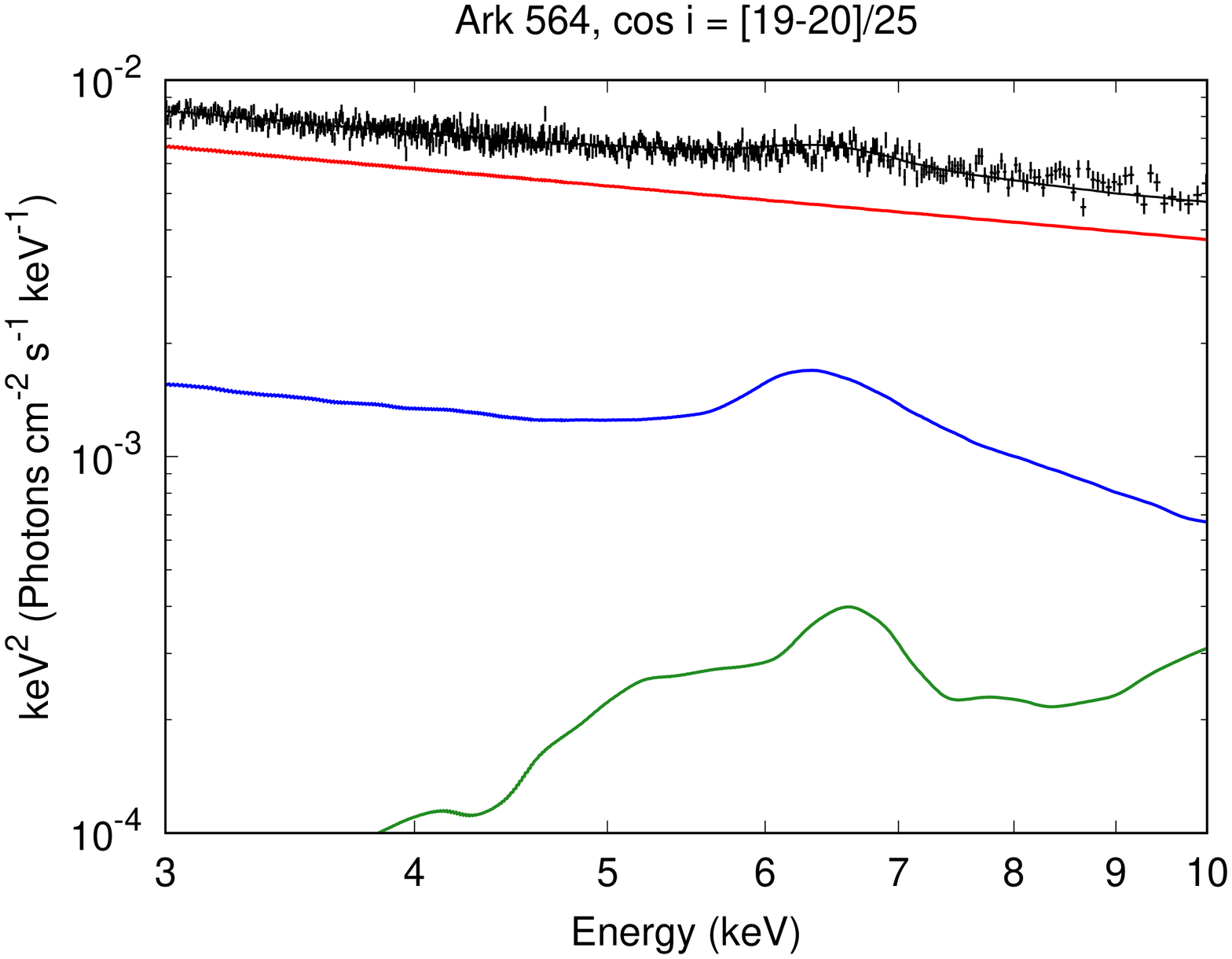}}
\resizebox{8.5cm}{!}{\includegraphics[width=2.7in,angle=270]{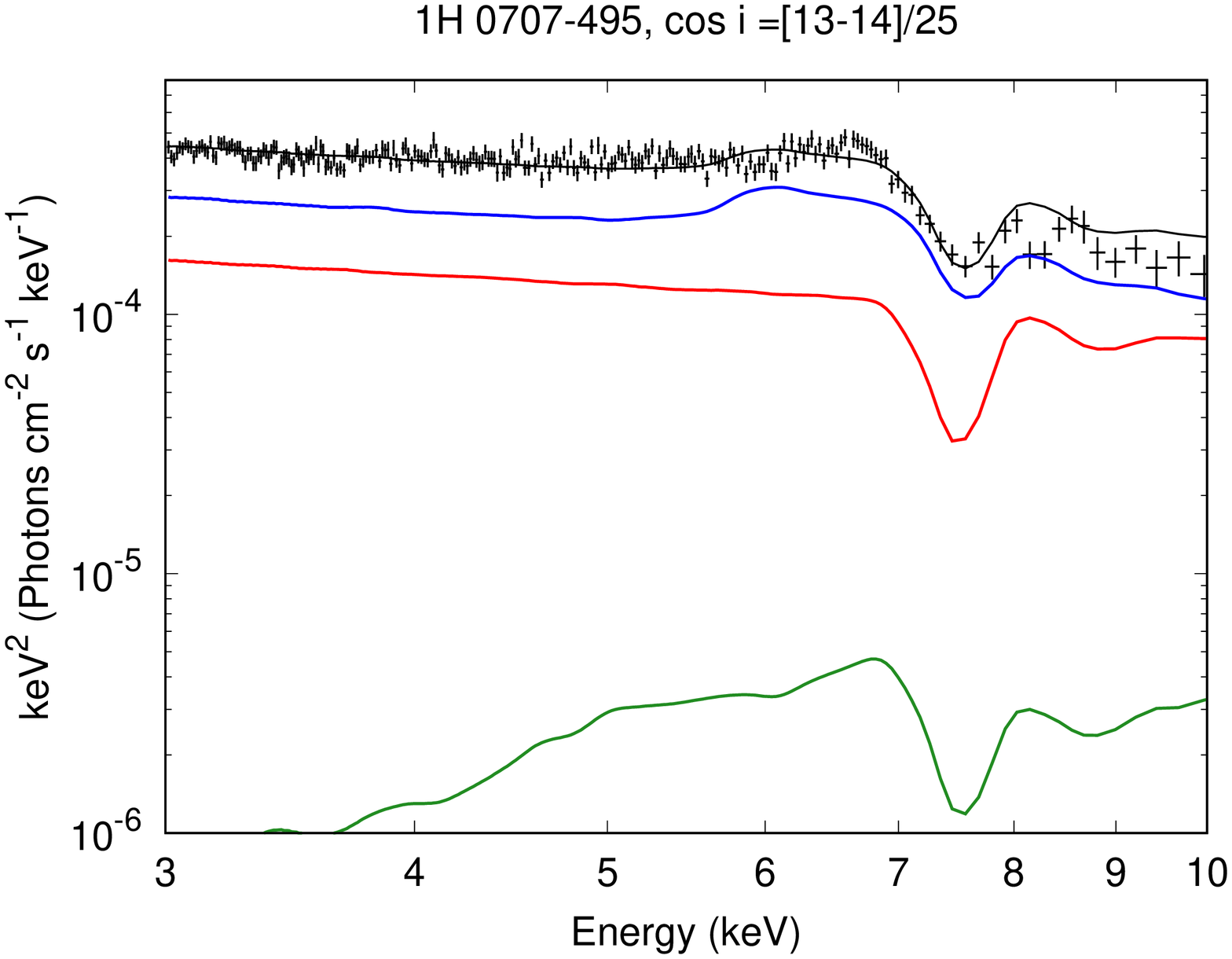}}
}
\subfigure{
\resizebox{8.5cm}{!}{\includegraphics[width=3.2in,angle=270]{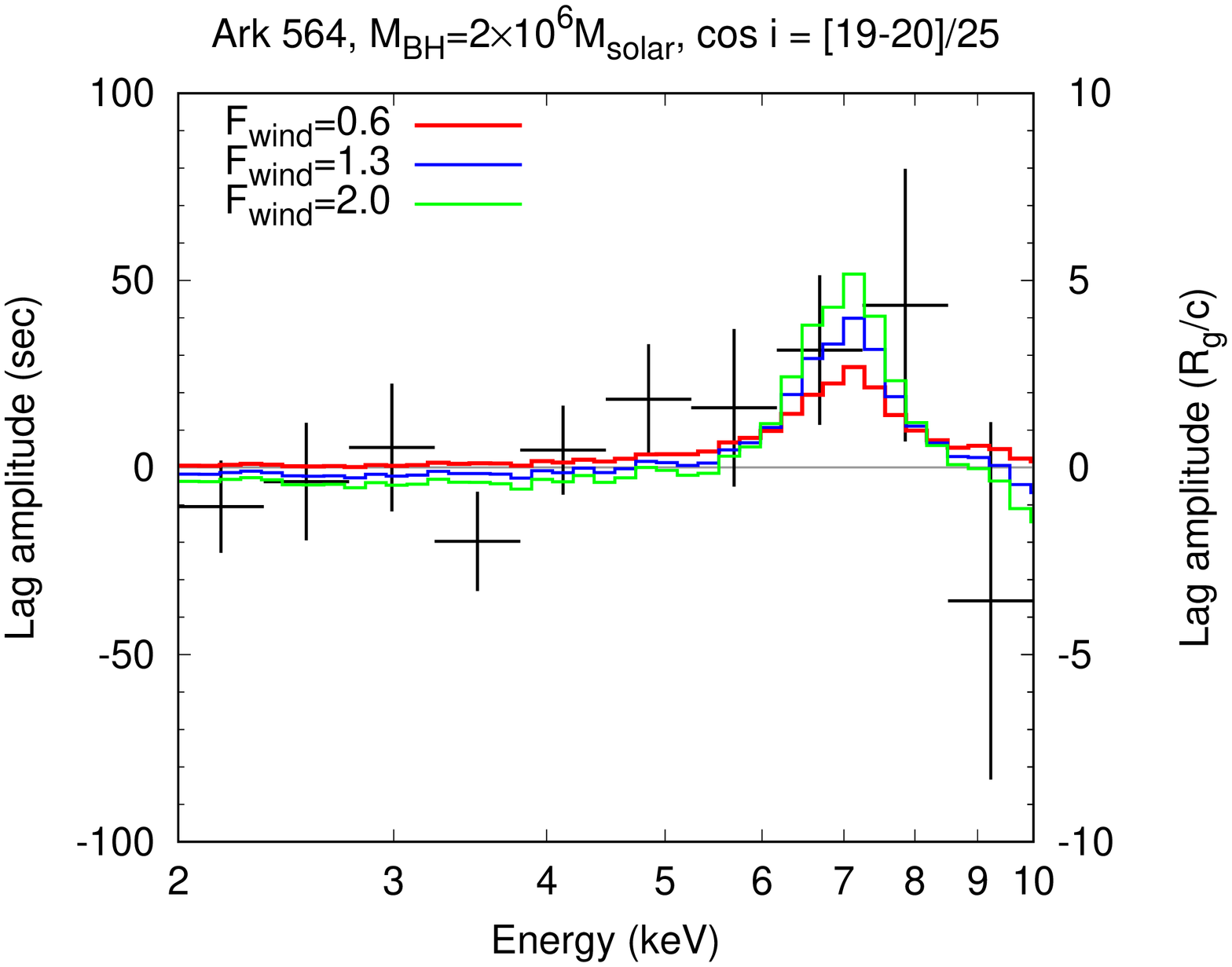}}
\resizebox{8.5cm}{!}{\includegraphics[width=3.2in,angle=270]{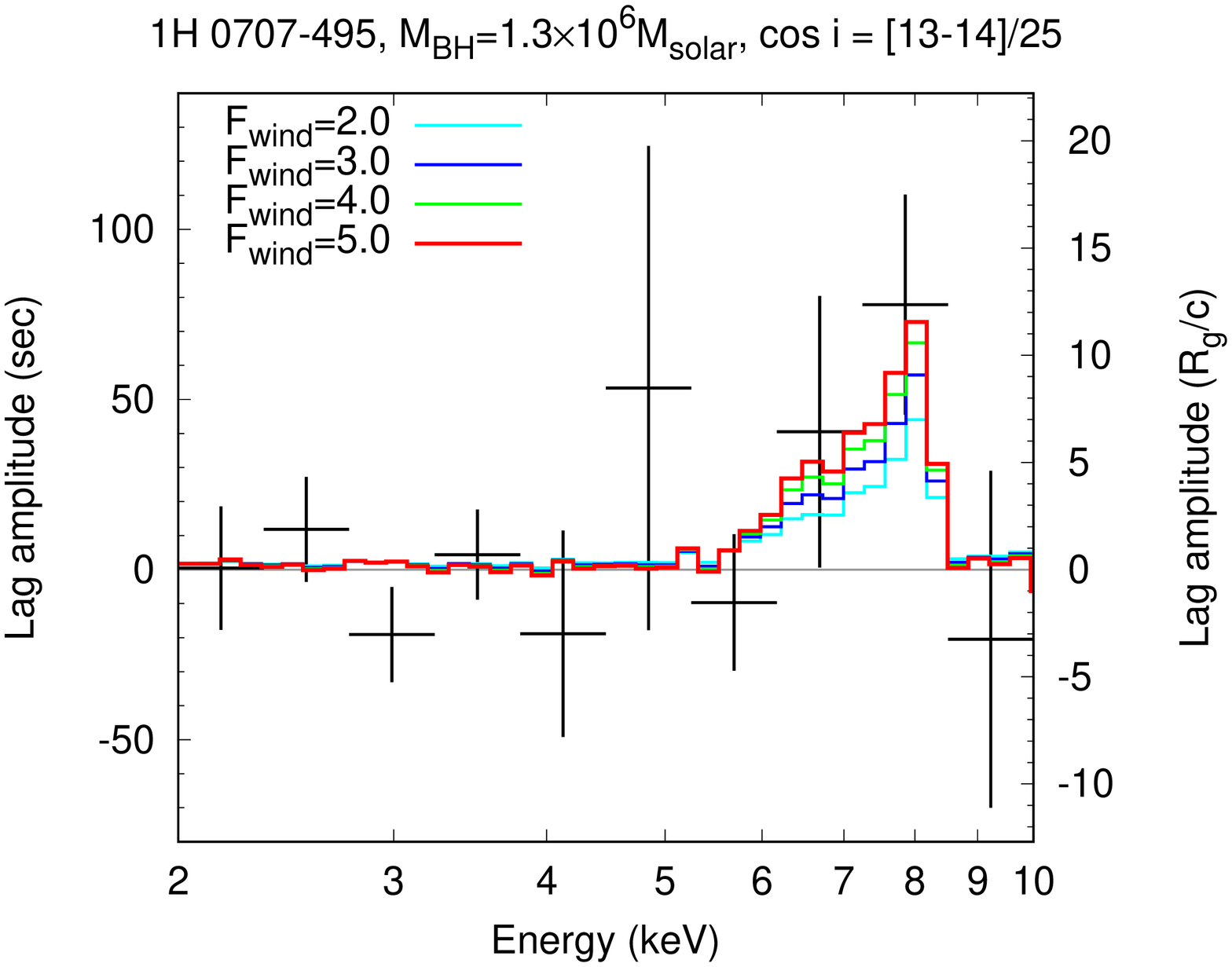}}
}
\subfigure{
\resizebox{8.5cm}{!}{\includegraphics[width=3.2in]{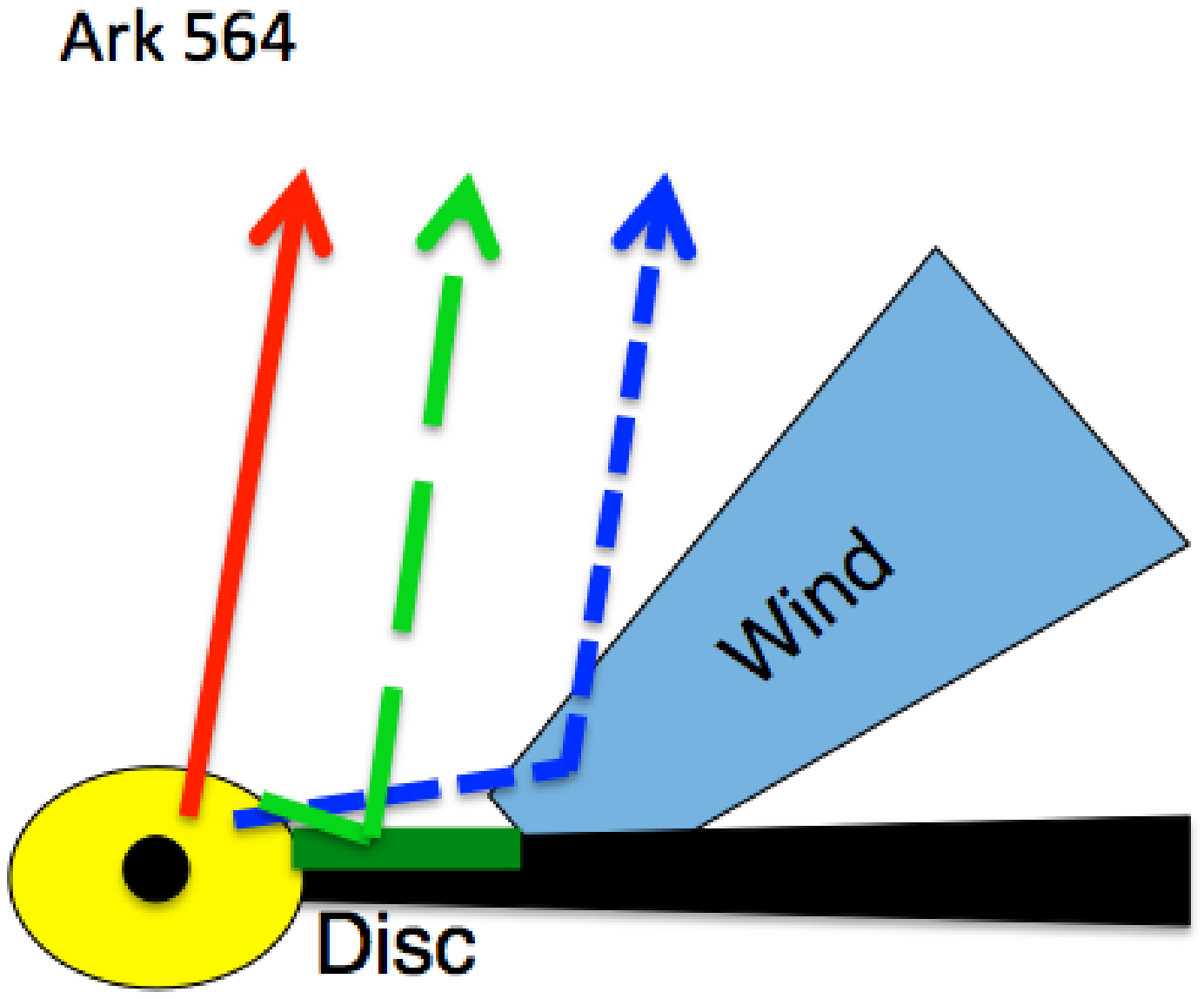}}
\resizebox{8.5cm}{!}{\includegraphics[width=3.2in]{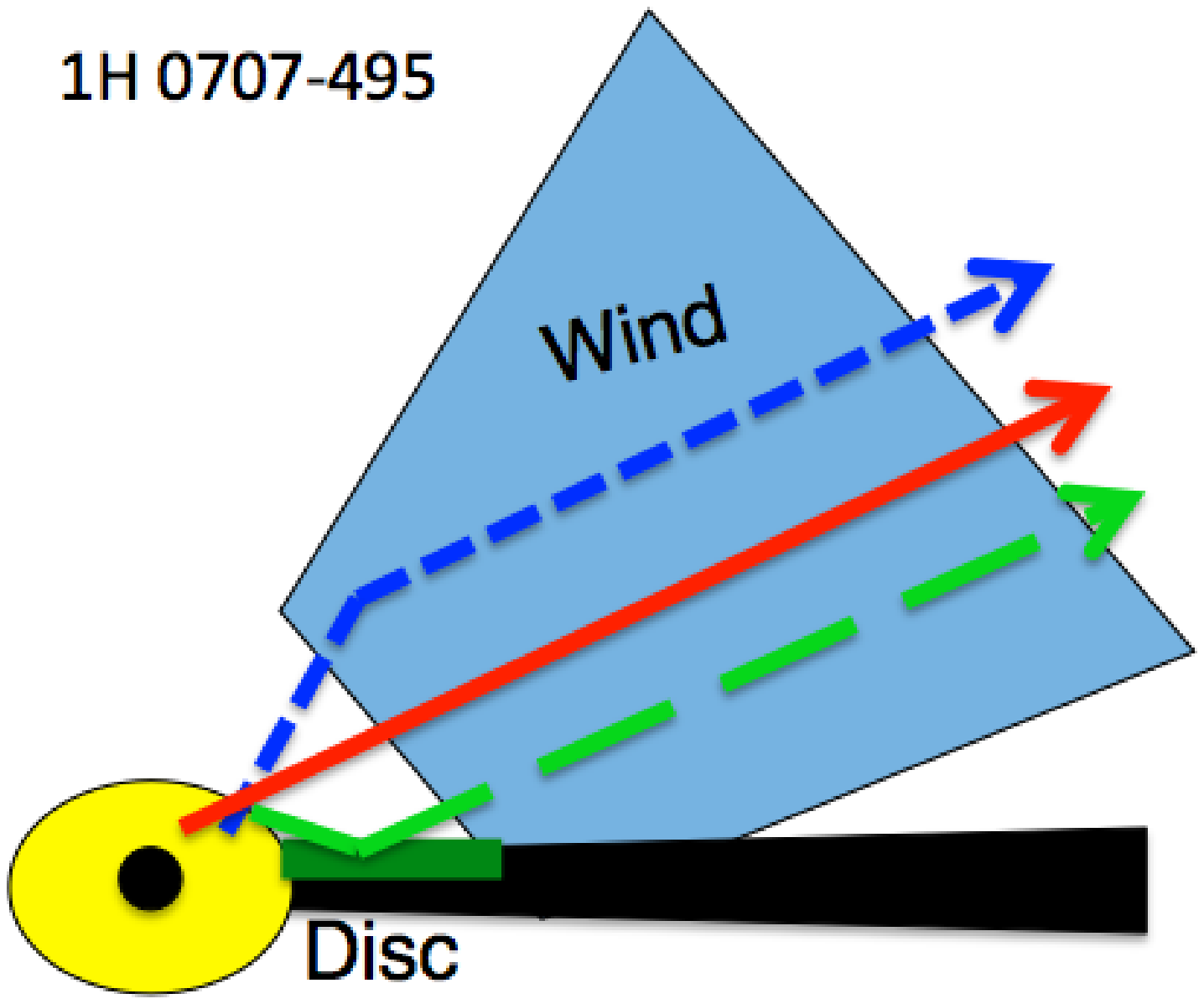}}
}
\caption{
(Upper) Energy spectral fitting for Ark 564 and 1H 0707--495. 
The symbols are same as figure \ref{fig:winddisc}(a) and the fitting parameters are shown in table \ref{tab:fitting}.
(Middle) Lag-energy plots of these sources at the frequency range of $c/250\,R_g-c/100\,R_g$. 
The red lines are the models determined by the spectral fitting in the upper panels, and
the different model lines show different $F_{\rm wind}$ values.
(Lower) Schematic pictures to explain the different energy spectra and the similar lag-energy spectra of the two targets. 
The extended X-ray corona (yellow) exists around the central BH. 
The red-solid, blue-dotted, and green-dashed arrows show the primary, wind-reflection, and disc-reflection components, respectively.
The wind solid angle is larger and the line of sight goes through the wind for 1H 0707--495,
whereas the solid angle is smaller and the line of sight does not go through the wind for Ark 564.
}
\label{fig:apply}
\end{center}
\end{figure*}

\subsection{Energy spectra}

We extract all the data for each source to calculate the time-averaged energy spectrum. 
In Ark 564, the variability is dominated by normalisation changes without significant spectral variations, 
so the time-averaged spectrum should give a good estimate of the instantaneous spectrum with better
statistics. 
On the other hand, 1H 0707--495 shows dramatic spectral variability as well as normalisation changes
(e.g. \citealt{miz14,hag16}), 
so the time-averaged spectrum may be different from the instantaneous spectra. 
These spectra are shown in the top panel of Fig.~\ref{fig:apply}. 
The lack of obvious absorption line in Ark 564 requires a face on geometry, 
while a P-Cygni-like line profile in 1H 0707--495 (e.g. \citealt{don07}) means that these data require a
line of sight intercepting the wind. 

We first explain the time-averaged energy spectra with our model (see Figs.\ \ref{fig:spectrum_wind} and \ref{fig:winddisc}a). 
We introduce two additional parameters to compensate for small differences from the specific model already calculated.
Firstly, we correct for small differences in spectral index compared to our simulation by multiplying the total spectrum by $E^{\Delta\Gamma}$, where $\Delta\Gamma=\Gamma_{\rm obs}-2.6$. 
Secondly, we allow the amount of wind-scattered emission to scale by a factor $F_{\rm wind}$ relative to the one in the model.
This assumes that the launch radius/velocity/density of the wind remains the same, but the solid angle and the total mass outflow rate in the wind can differ.  
For example, $F_{\rm wind}=1$ means that the flux ratio of the wind reflection component to the direct component is the same as the simulation, whereas
$F_{\rm wind}=2$ means that the ratio is twice as large as the one in the simulation.

The fitting results are shown in Tab.~\ref{tab:fitting} and the upper panels of Fig.~\ref{fig:apply}.
The best-fit inclination angles were obtained in the fitting from the 25  possible angle grid points in our calculation.
Ark 564 has no absorption features in its energy spectrum, and thus
requires fairly face-on inclination angles which do not intercept the wind. 
The $F_{\rm wind}$ value is 0.6, showing that the broad iron line in the data is slightly smaller than that in the original model. 
By contrast, 1H 0707--495 shows a clear P Cygni profile, requiring a high inclination angle which intercepts the wind material.
The best-fit $F_{\rm wind}$ is 5 times larger than that in our simulation, which implies a solid angle
$\Omega/2\pi=0.55\times 5 >1$ as viewed from a central source.
This indicates a rather geometrically-thick wind, as was independently suggested by the spectral fits in \citet{hag16}.

\begin{table}
	\centering
	\caption{Parameters of Ark 564 and 1H 0707--495}
	\label{tab:fitting}
	\begin{tabular}{ccc}
		\hline
		& Ark 564& 1H 0707--495 \\
		\hline
		inclination angle ($\cos i$) & [19--20]/25 & [13--14]/25  \\
              $\Delta\Gamma$ & 0.15 & 0.15 \\ 
		$F_{\rm wind}$ & 0.6 & 5 \\
		\hline
	\end{tabular}
\end{table}

\subsection{Lag-energy spectra}

As a next step, we investigate whether the lag features can be explained by the same parameters as those required by the energy spectra.
To do that, we need to know the mass of each AGN, because, as Fig.~\ref{fig:freqdepend} shows, 
the broad line in the model lag-energy spectra is not seen for frequencies higher than $\sim c/100\,R_g$, 
corresponding to the size scale of the wind.  
Although neither of these objects is included in the reverberation mapped AGN black hole mass database,
Ark 564 has some undersampled data with low variability in the `excluded datasets' category 
which gives an estimate of $\sim 10^6\,M_\odot$ \citep{she01,sha12}. 
Instead, the H$\beta$ line width gives $2.6\times 10^6\, M_\odot$ for Ark 564 \citep{bot04} and $2-4\times 10^6\, M_\odot$ in 1H 0707--495 \citep{don16}.

Here, we try instead to constrain their black hole masses from the observed X-ray reverberation lags.
We use the light curves with a bin-size of 100~s and calculated lags in each observation ID.
When the background flares interrupt the observation, we divided the light curve with several segments, calculated lags in each segment, and computed their average.
Finally we computed the average lag-energy spectra of all the observations (also see \citealt{miz18b}).
We compute the lag-energy spectra from the data over multiple frequency ranges,
$f_{\rm high}-f_{\rm low}$, where $f_{\rm high}/f_{\rm low}=10^{-0.4}$~Hz, and 
assume that the frequency at which the broad iron line lag diminishes corresponds to $c/100\,R_g$. 
For example in Ark 564, we calculated the lag-energy spectra in the frequency ranges of $10^{-[4.0-3.6]}$~Hz, $10^{-[3.9-3.5]}$~Hz, \dots, $10^{-[3.0-2.6]}$~Hz, 
and found that $10^{-[3.4-3.0]}$~Hz is the highest frequency range where the Fe-K lags are seen, i.e., the lag amplitude in the Fe-K energy bin exceeds the 0~s line with 1 sigma.
Similarly, in 1H 0707--495, the highest frequency at which the Fe-K lags are seen is $10^{-[3.2-2.8]}$~Hz. 
We associate these frequencies with those of $c/250\,R_g-c/100\,R_g$, and hence 
estimate the black home masses as $2\times10^6\,M_\odot$ and $1.3\times10^6\,M_\odot$ for Ark 564 and 1H 0707--495, respectively, consistent with those derived from X-ray variability (e.g. \citealt{pon12}). 

The middle panels of Fig.~\ref{fig:apply} show the lag-energy plots using the 2--10~keV lightcurve as the reference band, 
assuming the derived black hole masses and calculated over the frequency ranges identified above. 
The model lines are recalculated in the 2--10 keV band, with only the slight shift on the vertical axis from Fig.\ \ref{fig:lagvsE_wind} (in the 2--20 keV band).
Superimposed on this is the predicted lag-energy spectra using the $F_{\rm wind}$ values from
the best-fit spectral model for each source (red lines), 
as well as other $F_{\rm wind}$ values (green, blue, and cyan). 
We see that the lag-energy spectra predicted from the best-fit spectral model are consistent with the data. 
Consequently, both the energy spectra and the lag-energy plots are explained by the same wind parameters.

\section{Discussion}\label{sec5}
\subsection{Explanation of the difference between Ark 564 and 1H 0707--495}

In the previous section, we gave a quantitative model for the broad features 
in the Fe-K band of both the energy spectra and the lag spectra observed from Ark 564 and 1H 0707--495, in terms of the reprocessing in a disc wind.
In the framework of this model, the main differences between the two targets are the wind solid angle (as measured by $F_{\rm wind}$) and the inclination angle to our line of sight, as shown schematically in the  lower panel in Fig.~\ref{fig:apply}. 
The inclination angle is only a coincidence, but the difference in the wind solid angle implies a real difference between the two sources. 
We speculate that this is associated with the larger super-Eddington luminosity ($L/L_{\rm Edd}>10$; \citealt{don16}) in 1H 0707-495, than in Ark 564 which is only mildly super-Eddington ($L/L_{\rm Edd}\sim1$; \citealt{mul09}). 
The difference of the wind solid angle can be naturally explained by a larger radiation pressure at higher super-Eddington luminosities. 

\subsection{Complex absorption in the clumpy wind}

\begin{figure*}
\begin{center}
\includegraphics[width=6in]{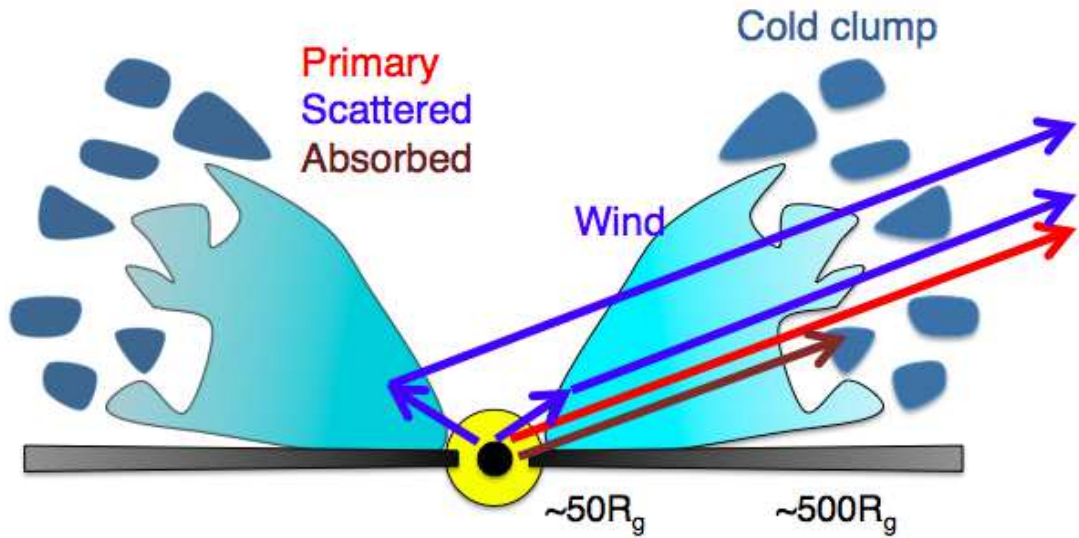}
\caption{
A physical picture to explain X-ray spectral variability in AGNs.
The yellow circle shows the extended X-ray emitting corona.
Disc winds are launched at $\sim 30-50R_g$.
The winds in the line of sight create blueshifted absorption lines as UFOs, and
those out of the line of sight within $\sim100\,R_g$ produce reverberation lags (blue).
The winds exhibit instability at $\gtrsim500\,R_g$ and make cold clumps.
They partially cover the corona and make a deep Fe-K edge (dark-red), which mimics the seemingly broad line feature.
Change of the partial covering fraction produces the deep rms dip in the Fe-K band.
Disc reflection is not shown here because it has minor contribution (see Fig. \ref{fig:winddisc}).
}
\label{fig:windpicture}
\end{center}
\end{figure*}

1H 0707--495 shows deep dips in its X-ray lightcurve,
where the drops are often as large as a factor 10 at $\sim 0.5$~keV 
while the high energy emission at 5--10~keV changes much less (e.g. \citealt{fab12}). 
The dip spectra are very faint, so they make little contribution to the time-averaged spectrum fit in Fig.~\ref{fig:apply}, 
but the dip spectral shape is so complex that cannot be explained by the hot and highly-ionised wind. 
Instead, the dip spectra can be explained by partial covering (in time and/or space) by almost neutral absorbers (e.g. \citealt{miz14,hag16,don16}).
\citet{don16} proposed that there are dense clumps embedded within the hot wind, 
perhaps produced as a result of thermal instability of the X-ray irradiated gas in pressure balance \citep{kro81}. 
However, the numerical simulations of the super-Eddington winds show clump formation driven instead by the Rayleigh-Taylor instability \citep{tak13,kob18}.
The radiation force exceeds the gravitational force so the heavier layer in the upper (inner) stream pushes the lighter one in the lower (outer) stream, which induces the instability, and 
produces cooler, denser (and hence less ionised) clumps in the outer region.  
Thus there will be two distinct absorption structures at different radii; 
the inner disc wind which is hot, highly-ionised and fairly smooth, and
then denser clumps in the outer wind ($\ge 500\, R_g$) which have $\tau\sim 1$ and a size of $\sim10\,R_g$.
 We sketch this geometry in Fig.~\ref{fig:windpicture}.  

Lines of sight which intersect the hot inner wind result in highly-ionised absorption lines (which may be observed as UFO), 
and also have a high probability to be occasionally and partially obscured by these denser clumps,
resulting in dramatic and complex absorption variability. 
Such variable partial covering models can fit the observed spectral variability (e.g. \citealt{miy12,miz14,iso16,yam16}). 
The occultation timescale is $t_{\rm occ}\sim2000 (M_{\rm BH}/2\times 10^6\,M_\odot) (R/500\,R_g)^{1/2} (D/10\,R_g)$~s \citep{miz17},
where $R$ and $D$ are the location and size of the cold clumps, respectively.
This timescale is very consistent with that of the observed deep (occultation-like) dips in the lightcurve of 1H 0707--495 (e.g. \citealt{miz14}).
Consequently, such complex absorption in the ``hot inner and clumpy outer wind model'' can explain the observational features of the super-Eddington sources.

\section{Conclusion} \label{sec6}

Short timescale lags around $\sim5\,R_g/c$ are widely seen over a broad Fe-K band in the lag-energy spectra of AGNs. 
These are most clearly seen in the super-Eddington sources Ark 564 and 1H 0707--495.
We show that these short lags are likely to be produced by a fast ($0.2c$) outflowing, highly-ionised wind at $50-100\,R_g$. 
The wind produces a broad emission line from combination of the rotation and the outflow, and
the few hundred $R_g$ size scale of the wind Fe-K reverberation lag is significantly diluted 
by the continuum emission underneath the iron line. 
We show that this model can simultaneously fit both the time-averaged energy spectra and the lag-energy spectra in Ark 564 and 1H 0707--495. 
These two sources show similar lag-energy spectra, but very different time average energy spectra, 
which we interpret as mainly due to difference in the inclination angle with respect to the wind. 
In Ark 564 our line of sight does not intercept the wind, while in 1H 0707--495 our line of sight does intercept the wind, resulting in a distinct blueshifted absorption line. 
There is also a noticeable difference that the broad iron emission line feature is stronger in 1H 0707--495, 
which we interpret as being due to a larger wind solid angle caused by the higher super-Eddington luminosity. 
Thus in Ark 564 we see the wind only via its small scattered emission, while
in 1H 0707--495 we see a prominent P-Cygni profile with the strong emission and absorption. 

While this ionised fast wind model can explain the short lags seen from the fast variability, 
it does not produce the dramatic spectral changes seen in 1H 0707--495 on longer timescales. 
Instead, this requires that there are denser and less-ionised clumps partially occulting the central X-ray source. 
Simulations of the super-Eddington winds show that such clumps are produced by Rayleigh-Taylor instabilities at $\sim 500\,R_g$. 
Consequently, the ``hot inner and clumpy outer wind model'' we propose here can explain all the observed features; 
the inner hot wind produces the Fe-K emission line seen in the time-averaged spectrum and the reverberation lags, 
together with the absorption line along the line of sight intersecting the wind,
whereas the outer cold clumps account for the longer term spectral variability as seen in the deep dips. 

Finally, we remark, as demonstrated in the present paper, that the reverberation provides us with a unique tool to constrain parameters of the material outside of the line of sight. 
Estimating the solid angle subtended by the wind is crucial to explore energetics of the AGN outflows, and hence to quantify their contribution to the AGN feedback (see e.g. \citealt{kin15}). 

\section*{Acknowledgements}
Authors are financially supported by the JSPS/MEXT KAKENHI Grant Numbers JP15J07567 (MM), JP16K05309 (KE), and JP24105007, JP15H03642, JP16K05309 (MT).
MM acknowledges JSPS overseas research fellowship.
CD acknowledges the Science and Technology Facilities Council (STFC)
through grant ST/P000541/1 for support.

%The Acknowledgements section is not numbered. Here you can thank helpful
%colleagues, acknowledge funding agencies, telescopes and facilities used etc.

%%%%%%%%%%%%%%%%%%%%%%%%%%%%%%%%%%%%%%%%%%%%%%%%%%



%\bibliographystyle{mnras}


\bibliographystyle{mnras}
\bibliography{mn-jour,00}
%%%%%%%%%%%%%%%%%%%%%%%%%%%%%%%%%%?%%%%%%%%%%%%%%%%

%%%%%%%%%%%%%%%%% APPENDICES %%%%%%%%%%%%%%%%%%%%%

%\appendix
%\input{app}



%%%%%%%%%%%%%%%%%%%%%%%%%%%%%%%%%%%%%%%%%%%%%%%%%%

% Don't change these lines
\bsp	% typesetting comment
\label{lastpage}
\end{document}